\documentclass[pdftex,twocolumn,epjc3]{svjour3}          

\RequirePackage[T1]{fontenc}

\smartqed  

\RequirePackage{graphicx}
\RequirePackage{mathptmx}      
\RequirePackage{flushend}
\RequirePackage[numbers,sort&compress]{natbib}
\RequirePackage[colorlinks,citecolor=blue,urlcolor=blue,linkcolor=blue]{hyperref}

\journalname{Eur. Phys. J. C}

\begin{document}

\title{Cold Nuclear Matter Effects on Dijet Productions in Relativistic Heavy-ion Reactions at LHC 
}


\author{Yuncun He\thanksref{addr1,e1}
        \and
        Ben-Wei Zhang\thanksref{addr1,addr2,e2} 
         \and
        Enke Wang\thanksref{addr1,addr2,e3} 
}

\thankstext{e1}{e-mail: heyc@iopp.ccnu.edu.cn}
\thankstext{e2}{e-mail: bwzhang@iopp.ccnu.edu.cn}
\thankstext{e3}{e-mail: wangek@iopp.ccnu.edu.cn}

\institute{Institute of Particle Physics, Central China Normal University, Wuhan 430079, China\label{addr1}
          \and
          Key Laboratory of Quark \& Lepton Physics (Central China Normal University), Ministry of Education, China\label{addr2}
}

\date{Received: date / Accepted: date}

\maketitle

\begin{abstract}
 We investigate the cold nuclear matter(CNM) effects on dijet productions  in high-energy nuclear collisions at LHC with the
 next-to-leading order perturbative QCD. The nuclear modifications for dijet angular distributions, dijet invariant mass spectra, dijet transverse momentum spectra and  dijet momentum imbalance due to CNM effects are calculated by incorporating EPS, EKS, HKN and
 DS param-etrization sets of parton distributions in nucleus . It is found that dijet angular distributions and dijet momentum imbalance are insensitive to the initial-state CNM effects and thus provide optimal tools to study the final-state hot QGP effects such as jet quenching. On the other hand,  the invariant mass spectra and the transverse momentum spectra of dijet are generally enhanced in a wide region of the invariant mass or transverse momentum due to CNM effects with a feature opposite to the expected suppression because of the final-state parton energy loss effect in the QGP. The difference of EPS, EKS, HKN and DS parametrization sets of nuclear parton distribution functions is appreciable for dijet invariant mass spectra and transverse momentum spectra at p+Pb collisions,  and becomes more pronounced for those at Pb+Pb reactions.
\end{abstract}

\section{Introduction}
In relativistic heavy-ion collisions with a large amount of energies deposited in the collision center quarks and gluons confined in nucleons should be liberated to form
a new kind of matter -- quark gluon plasma (QGP). Hard processes with large momentum transfer have long been regarded as good probes of the QGP created in
high-energy nuclear reactions~\cite{HP2010} and an important probe among them is the jet quenching, or the energy loss of an energetic parton in QCD matter~\cite{Wang:1991xy, Baier:1996sk, Gyulassy:2000fs,Wang:2001ifa}. With the upgrading facilities
at RHIC and
especially the unprecedented colliding energies at LHC, measurements of leading particle production at large transverse momentum have been extended to reconstructed jet
production,  and several theoretical studies of jet production in high-energy nuclear collisions have been made by investigating hadron multiplicity inside a jet~\cite{Borghini:2005em}, jet shape~\cite{Lokhtin:2006dp,Vitev:2008rz}, inclusive jet cross section at the next-to-leading order (NLO)~\cite{Vitev:2009rd}, and $Z^0/\gamma ^*$-tagged jets~\cite{Neufeld:2010fj}.
Recently ATLAS and CMS have made the first successful measurement of jet production in A+A reactions at LHC by observing dijet asymmetry in Lead-Lead collisions~\cite{Aad:2010bu,Chatrchyan:2011sx}, which has aroused
intense theoretical investigations~\cite{Qin:2010mn,Lokhtin:2011qq,Young:2011qx,He:2011pd}. It has been demonstrated that jet quenching, or final-state parton energy loss in the QGP should be responsible for the large
dijet momentum imbalance in nuclear reactions at LHC~\cite{CasalderreySolana:2010eh}
with some caution of the background fluctuation effect in heavy-ion collisions~\cite{Cacciari:2011tm}.

In heavy-ion collisions because the initial-state colliding objects are nuclei instead of nucleons, cold nuclear matter (CNM) effects such as shadowing effects,
anti-shadowing effects, EMC effects {\it etc.} will manifest themselves in hard processes, and it has been shown that CNM effects will play an important role in
productions of leading hadrons~\cite{Neufeld:2010dz}, direct photons~\cite{Vitev:2008vk,Zhou:2010zzm, Arleo:2011gc}, Drell-Yan dilepton~\cite{Xing:2011fb,Johnson:2000ph,Neufeld:2010dz}, heavy flavor mesons~\cite{Ferreiro:2008wc,Sharma:2009hn,Duan:2011gu} as well as inclusive jets~\cite{Vitev:2009rd,Zhang:2011ak} and other hard processes in nuclear collisions. It will be of great interest to see how large
CNM effects will be in dijet production of high-energy nucleus collisions, which will then help us understand the final-state jet quenching effect more precisely.
It has been known that for hard processes in nuclear collisions
with the assumption of the leading twist factorization
of perturbative QCD (pQCD), the main CNM effects can be phenomenologically parameterized the difference between parton distribution functions (PDFs) in nucleon and
nuclear parton distribution functions (nPDFs)~\cite{Eskola:2009uj}. Other theoretical models have also been developed to pinpoint the underlining mechanisms of different CNM effects with
many-body QCD and related phenomenologies~\cite{Frankfurt:2011cs, Vitev:2007ve}. In this paper we will investigate CNM effects on dijet production in p+Pb and Pb+Pb collisions at LHC at the NLO perturbative QCD by incorporating
4 parametrization sets of nPDFs -- EPS~\cite{Eskola:2009uj}, EKS~\cite{Eskola:1998df}, HKN~\cite{Hirai:2007sx} and
DS~\cite{deFlorian:2003qf}. Four
dijet observables--- dijet angular distributions, dijet invariant mass spectra, dijet transverse momentum spectra and
 dijet momentum imbalance with CNM effects in p+Pb and Pb+Pb collisions at LHC will be calculated as well as the
 nuclear modification factors for these dijet observables due to CNM effects.

Our work is organized as follows: in Section 2, we give the formalism of dijet productions in nucleon-nucleon collisions
with perturbative QCD. Dijet observables for p+Pb and Pb+Pb collisions  with CNM effects at LHC by utilizing 4 different sets of nPDFs
and their physics implications are studied in Section 3.  At last we give a brief summary in Section 4.

\section{Dijet Production in nucleon-nucleon Collisions}
 Dijet productions in elementary nucleon-nucleon collisions provide the base line for dijet productions in
  relativistic heavy-ion collisions. With the perturbative QCD factorization approach~\cite{Owens:1986mp, Campbell:2006wx},
 we can obtain leading-order (LO) invariant transverse momentum spectrum for
 dijet in elementary collisions
\begin{eqnarray}
   \frac{d\sigma}{dy_1dy_2dE^2_T}=\sum_{abcd}x_af_{a/A}(x_a)x_bf_{b/B}(x_b)\frac{d\sigma}{dt}(ab\rightarrow cd)
   \label{di-pt},
\end{eqnarray}
where $f_{a,b}(x_{a,b})$ is the parton distribution function in nucleon, and $x_a$, $x_b$ are the momentum fractions carried
by participating partons from nucleons. They have relationships with transverse energy $E_T$ and rapidity $y_1$, $y_2$ of final particles as
\begin{eqnarray}
   x_{a}=\frac{E_T}{\sqrt{s}}(e^{y_1}+e^{y_2}),
   x_{b}=\frac{E_T}{\sqrt{s}}(e^{-y_1}+e^{-y_2})
   \label{x}.
\end{eqnarray}
$d\sigma/dt$ is the elementary scattering cross section for partons at tree level. And $t$ is the Mandelstam variable.

 At next-to-leading order (NLO), jets can be defined by jet-finding algorithms with a radius parameter in rapidity $y$
 and azimuthal angle $\phi$ plane, $R=\sqrt{\Delta y^2+\Delta \phi^2}$ ~\cite{Ellis:2007ib}. The cross section of dijet production at NLO can be expressed as~\cite{Kunszt:1992tn},
\begin{eqnarray}
   \frac{d\sigma}{dV_J}&=&\frac{1}{2!}\int dy_1dE_{T2}dy_2d\phi_2 \frac{d\sigma(2\rightarrow2)}{dy_1dE_{T2}dy_2d\phi_2}S_2(p_1^{\mu},p_2^{\mu})\nonumber \\
   &~&+\frac{1}{3!}\int dy_1dE_{T2}dy_2d\phi_2dE_{T3}dy_3d\phi_3 \nonumber \\ &~&\times\frac{d\sigma(2\rightarrow3)}{dy_1dE_{T2}dy_2d\phi_2E_{T3}dy_3d\phi_3}S_3(p_1^{\mu},p_2^{\mu},p_3^{\mu})
   \label{di-pt1}.
\end{eqnarray}
Here, $V_{J}$ represents the physical quantity of final state, and $p_i^{\mu}$, $E_{T\, i}$, $y_i$, $\phi_i$ are the four-momentum, transverse energy, rapidity, azimuthal angle of the $i-$th($i=$~1, 2, 3) particle, respectively. The first term on the right-hand denotes the contribution from $2\rightarrow2$ processes including NLO virtual corrections. The second term is the contribution from $2\rightarrow3$ processes. And the functions $S$ contain the jet finding algorithm.  In this paper we will estimate $V_{J}$ as angular, invariant mass and final transverse momentum of dijet by utilizing the EKS framework of a NLO calculation of jet productions in hadron-hadron collisions
~\cite{Kunszt:1992tn,Ellis:1990ek,Ellis:2007ib}.

\begin{figure}
\includegraphics[angle=90,width=85mm]{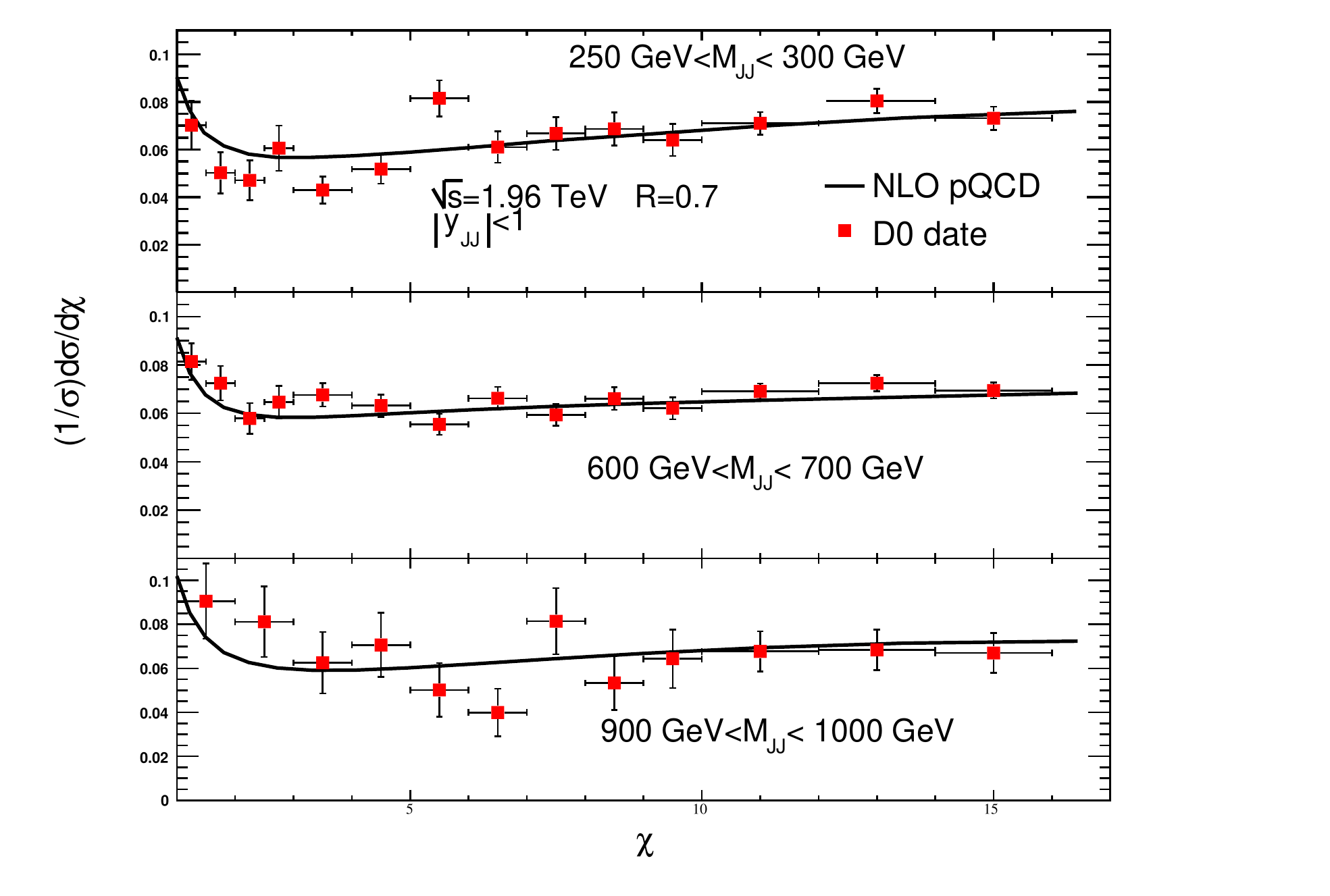}
\includegraphics[width=85mm]{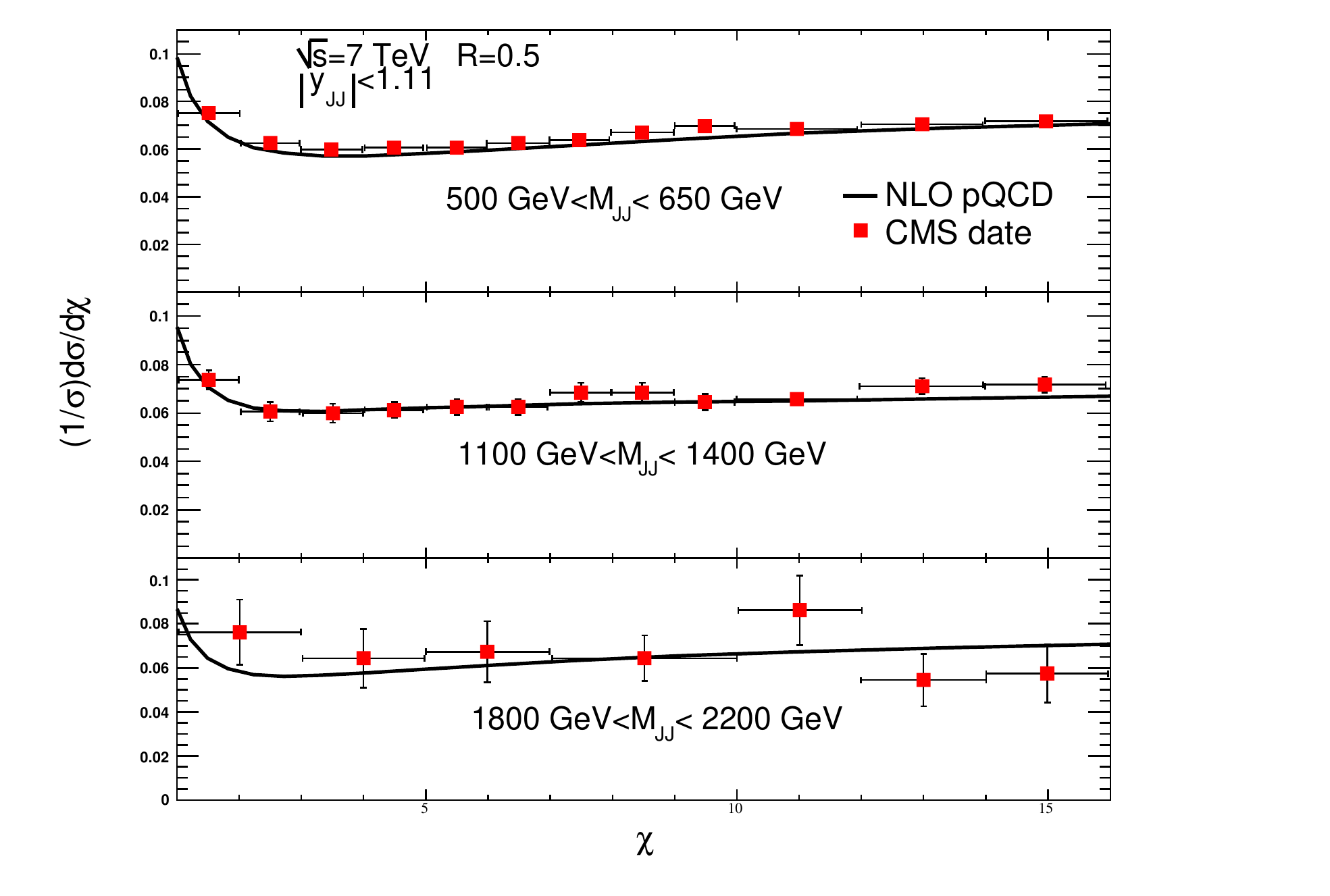}
    \caption{The angular distributions of dijet are evaluated at NLO and compared to experimental results form D0 collaboration in p+$\bar{p}$ collisions at $\sqrt{s}=1.96$~TeV ~\cite{:2009mh} and CMS collaboration in p+p collisions at $\sqrt{s}=7$~ TeV~\cite{Khachatryan:2011as}.}
    \label{1.dchi}
\end{figure}

The rapidity of a dijet system can be defined as
\begin{eqnarray}
   y_{JJ}=\frac{y_{1}+y_{2}}{2},
   y^{*}=\frac{|y_{1}-y_{2}|}{2},  \label{y}
\end{eqnarray}
and the angular of system is related to rapidity as $\chi=e^{2y^{*}}$. For massless partons, $\chi$ can be written as
\begin{eqnarray}
   \chi=\frac{1+\cos\theta^{*}}{1-\cos\theta^{*}},
\end{eqnarray}
where $\theta^{*}$ is the polar scattering angle of the outgoing jets in the dijet center-of-mass frame~\cite{theta}.

We can also define the dijet invariant mass $M_{JJ}$ as the invariant mass $[(\sum p_n^{\mu})^2]^{1/2}$ of all particles in the two jets. At LO, it has a form
\begin{eqnarray}
   M_{JJ}^2=2E^2_T[1+\cosh(y_1-y_2)] \, .
   \label{M}
\end{eqnarray}
The invariant mass cross section of dijet as well as dijet angular distribution has provided a good tool to test predictions of perturbative QCD and predictions beyond the Standard Model(SM) such as quark compositeness, extra spatial dimensions~\cite{Eichten:1983hw,Eichten:1984eu,Lane:1996gr,ArkaniHamed:1998rs,Atwood:1999qd,Dienes:1998vg,Pomarol:1998sd,Cheung:2001mq}.

\begin{figure}
\includegraphics[width=85mm]{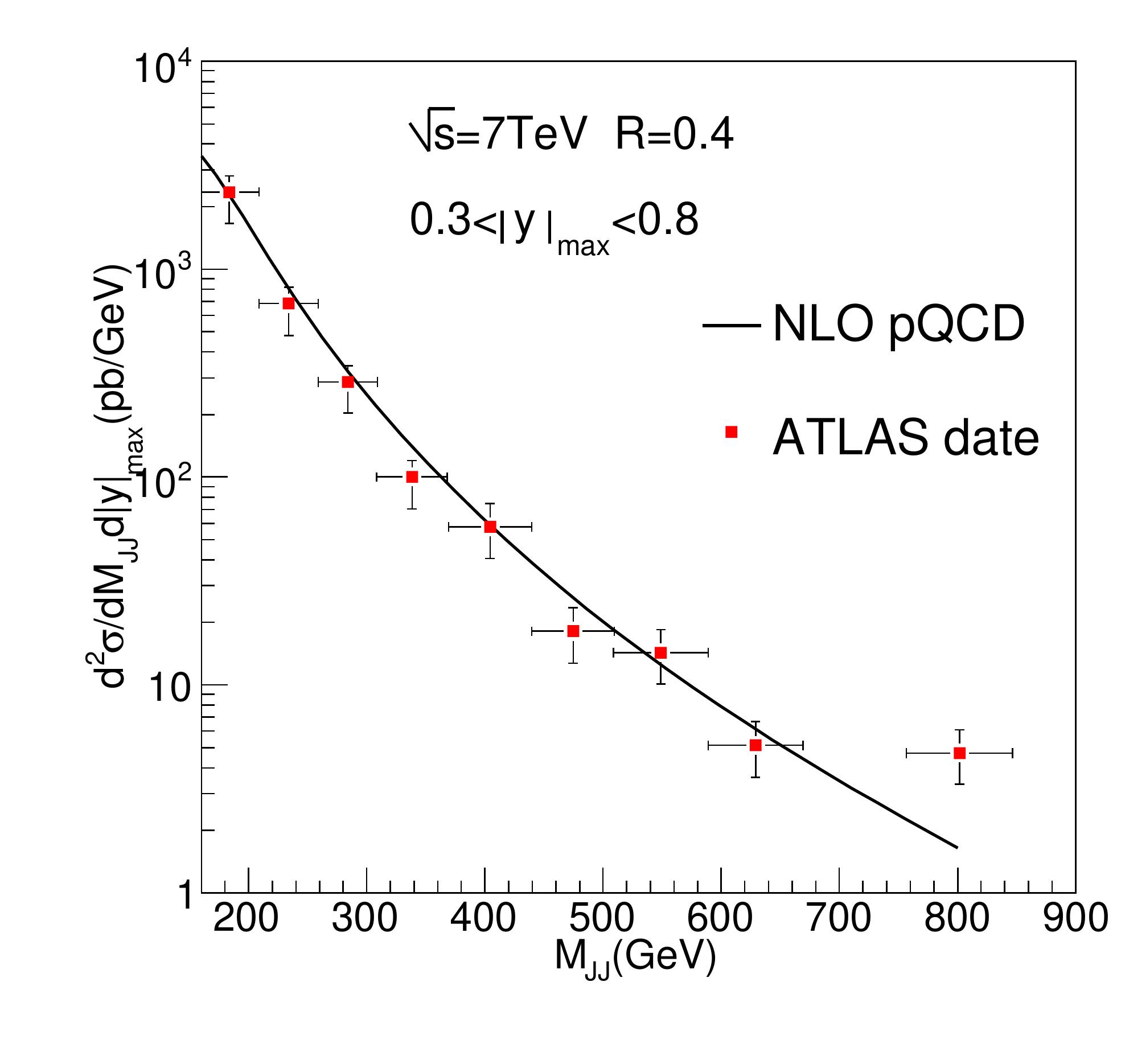}
    \caption{Comparison between dijet invariant mass spectrum at the NLO with the jet size $R=0.4$ and ATLAS data in p+p collisions at $\sqrt{s}=7$~ TeV~\cite{:2010wv}. }
    \label{2.dmjj}
\end{figure}

Our results for angular distributions at NLO shown in Figure~\ref{1.dchi} are compared to data from D0 collaboration and CMS collaboration with the utilizing of the CTEQ6.1M parton distribution functions (PDFs)~\cite{Pumplin:2002vw}. D0 detector in Run \uppercase\expandafter{\romannumeral2} of the Fermilab Tevatron Collider measured the angular distribution with the jet size $R=0.7$, $|y_{JJ}|<1$ at $\sqrt{s}=1.96$~TeV~\cite{:2009mh}. And the CMS collaboration measured that with the jet size $R=0.5$, $|y_{JJ}|<1.11$ at $\sqrt{s}=7$~TeV~\cite{Khachatryan:2011as}. We confront the
 theoretical results in p+$\bar{p}$ collisions with D0 measurements in the invariant mass intervals of 250~GeV$<M_{JJ}<300$~GeV , 600~GeV$<M_{JJ}<700$~GeV, 900~GeV$<M_{JJ}<1000$~GeV, and the numerical simulations in p+p reactions with CMS in the invariant mass intervals of
 500~GeV $<M_{JJ}< 650$~GeV , 1100~GeV$<M_{JJ}<1400$~GeV, 1800~GeV$<M_{JJ}<2200$~GeV.
 We observe a good agreement between theory and experimental data in different intervals of invariant mass. The most symmetrical production at $y_1=y_2$, namely $\chi=1$ is the largest. When the scattering angle $\theta^{*}$ is very small, the angular distribution is proportional to the Rutherford cross section~\cite{D0 PRD}. Hereby, the dijet angular distribution approaches a constant at large $\chi$.

\begin{figure}
\includegraphics[angle=90,width=85mm]{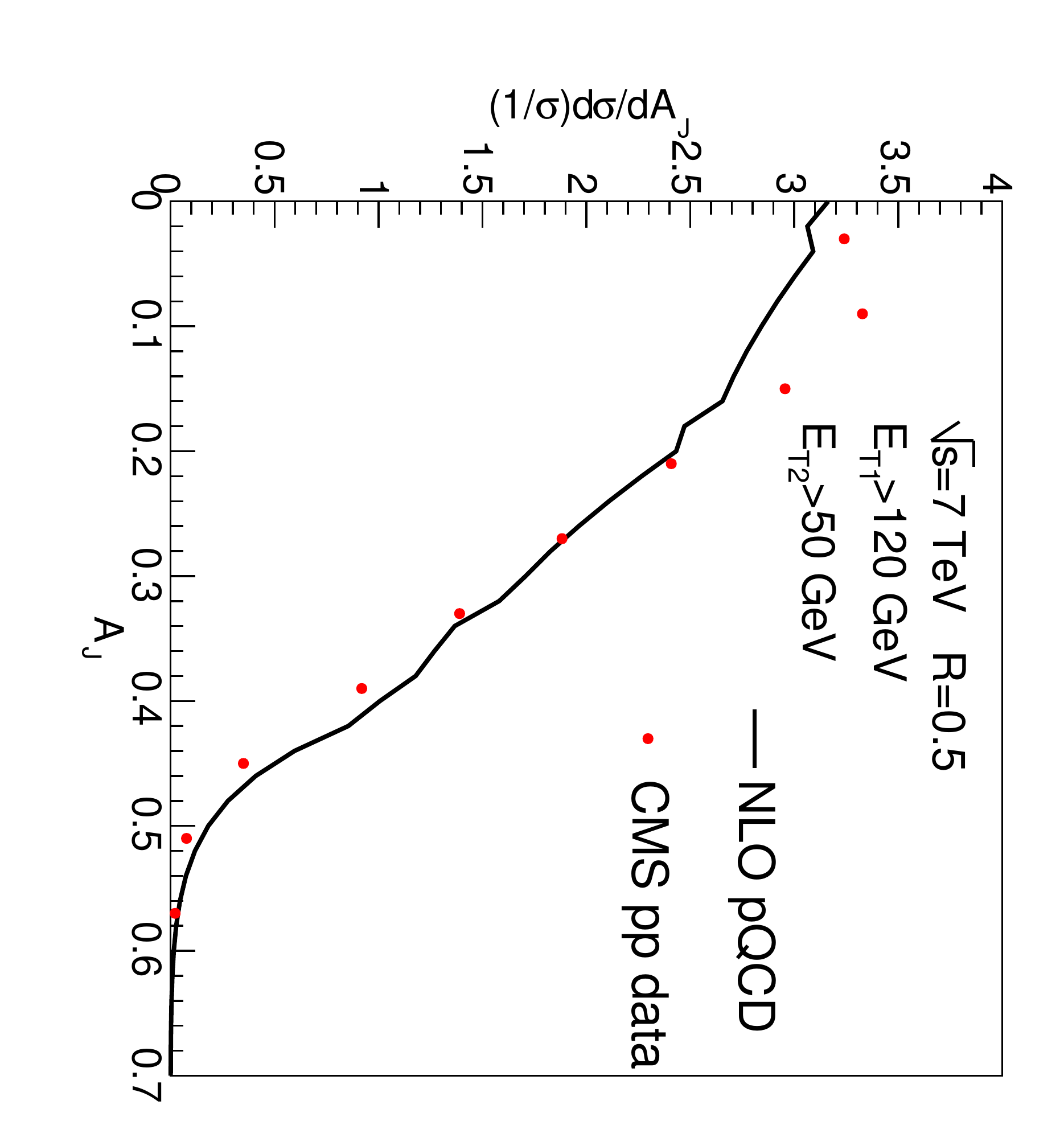}
    \caption{Comparison between dijet asymmetry distribution at the NLO with the jet size $R=0.5$ and the CMS measurement in p+p collisions at $\sqrt{s}=7$~TeV~\cite{Chatrchyan:2011sx} .}
    \label{daj}
\end{figure}

The ATLAS collaboration at the LHC has published recent results about the invariant mass spectrum of dijet in the bin of maximum rapidity of the two leading jets, $|y|_{max}$
=max$(|y_1|,|y_2|)$ in p+p collisions at  $\sqrt{s}=7$~TeV~\cite{:2010wv}. We test the dijet invariant mass cross
section in the rapidity bin of $0.3<|y|_{max}<0.8$ with the jet size $R=0.4$ in Figure~\ref{2.dmjj}. The NLO pQCD theory can describe the differential yield well in the mass range from 200~GeV to 650~GeV where the cross section falls by more than two orders of magnitude .

At the NLO the exact transverse momentum balance between $E_{T\, 1}$ and $ E_{T\, 2}$ should be broken
because of additional gluon radiation in $2 \rightarrow 3$ processes.
To quantify the transverse momentum
imbalance between the leading jet and the subleading jet one defines the dijet transverse momentum asymmetry
$A_J$ as,
\begin{equation}
A_J=\frac{E_{T\,1}-E_{T\, 2}}{E_{T\,1}+E_{T\, 2}} \;.
\label{ajdef}
\end{equation}
The $A_J$ distribution can be expressed from the dijet transverse energy spectrum as follows~\cite{He:2011pd}:
\begin{eqnarray}
\frac{d\sigma}{dA_J}&=& \int_{y_{1\,\min}}^{y_{1\,\max}} \int_{y_{2\,\min}}^{y_{2\,\max}}
\int_{E_{T\,2\, \min}}^{E_{T\,1}} dE_{T\, 2}  \; \frac{2 E_{T\,2}}{(1-A_J)^2} \nonumber \\[2ex]
&& \times\frac{d\sigma[E_{T\,1}(A_J,E_{T\,2})]}{dy_1dy_2 dE_{T\,1}dE_{T\,2}} \;,
\label{ajcalc}
\end{eqnarray}
where $E_{T\,1} = E_{T\,2}(1+A_J)/(1-A_J)$ is the transverse energy for leading jet. At the LO we always have $A_J = 0$ due to the equal transverse energies of two jets.

The CMS collaboration has carried out the asymmetry distribution of dijet in p+p collisions at $\sqrt{s}=7$~TeV~\cite{Chatrchyan:2011sx}, requiring the leading jet with the transverse energy of  $E_{T1}>120$~GeV, and the subleading jet with $E_{T2}>50$~GeV in opposite $\Delta \phi=|\phi_1-\phi_2|>2\pi/3$. The numerical result from Eq.~(\ref{ajcalc}) is given in Figure~\ref{daj} and confronted with the experimental data from CMS. The result from the NLO pQCD describes the CMS dijet asymmetry data very well.

\section{Dijet production in heavy-ion collisions with CNM effects}
Extrapolating the cross sections in p+p collisions to p+A and A+A, one can replace $f_{a/A}$ with different sets of
nPDFs ( EPS09~\cite{Eskola:2009uj}, EKS98~\cite{Eskola:1998df}, HKN~\cite{Hirai:2007sx}, DS~\cite{deFlorian:2003qf} ).
They modify the parton distribution in nucleon with shadowing (anti-shadowing) factors which reflect the modification
for probability of initial participating partons found in the nucleus. At the same time, isospin symmetry for bound
 protons and neutrons, EMC effect and Fermi motion in nucleus are also included. For a nucleus $A$ with $Z$ protons, the factorization form is
\begin{eqnarray}
   f_{a/A}(x,Q^2)&=&R_{a/A}(x,Q^2)[\frac{Z}{A}f_{a/p}(x,Q^2)\nonumber \\
   &~&+(1-\frac{Z}{A})f_{a/n}(x,Q^2)]
   \label{Rchi}.
\end{eqnarray}
To make the NLO computations of dijet in relativistic heavy-ion collisions we combine the NLO calculation of
dijets in hadron-hadron collisions
in Eq.~({\ref{di-pt1}}) with the nPDFs of EPS09 NLO, HKN NLO  and DS NLO sets,
respectively. Because EKS98 does not have NLO nPDFs the LO parametrization of EKS98 nPDFs will be used when calculations with EKS are carried out.

We also need insert thickness function $t_A(\mathbf{b})$ in the whole reaction
plane, which indicates the number of participating nucleons for unit area in the nucleus . With the Glauber model the spectrum for the dijet production in p+A collisions is given by
\begin{eqnarray}
   \frac{d\sigma_{pA}}{dy_1dy_2dE^2_T}&=&\sum_{abcd}\int d^{2} \mathbf{b}t_A(\mathbf{b})x_af_{a/p}(x_a)x_bf_{b/A}(x_b)\nonumber \\
   &~&\times\frac{d\sigma}{dt}(ab\rightarrow cd)
   \label{di-pt},
\end{eqnarray}
and for $A+A$ collisions, the spectrum is
\begin{eqnarray}
   \frac{d\sigma_{AA}}{dy_1dy_2dE^2_T}&=&\sum_{abcd}\int d^{2}\mathbf{b}\int d^{2}\mathbf{r}t_A(\mathbf{r})t_B(\mathbf{b}-\mathbf{r})\nonumber \\
   &~&\times x_af_{a/A}(x_a)x_bf_{b/B}(x_b)\frac{d\sigma}{dt}(ab\rightarrow cd)
   \label{di-pt}.
\end{eqnarray}

\begin{figure}
\includegraphics[width=85mm]{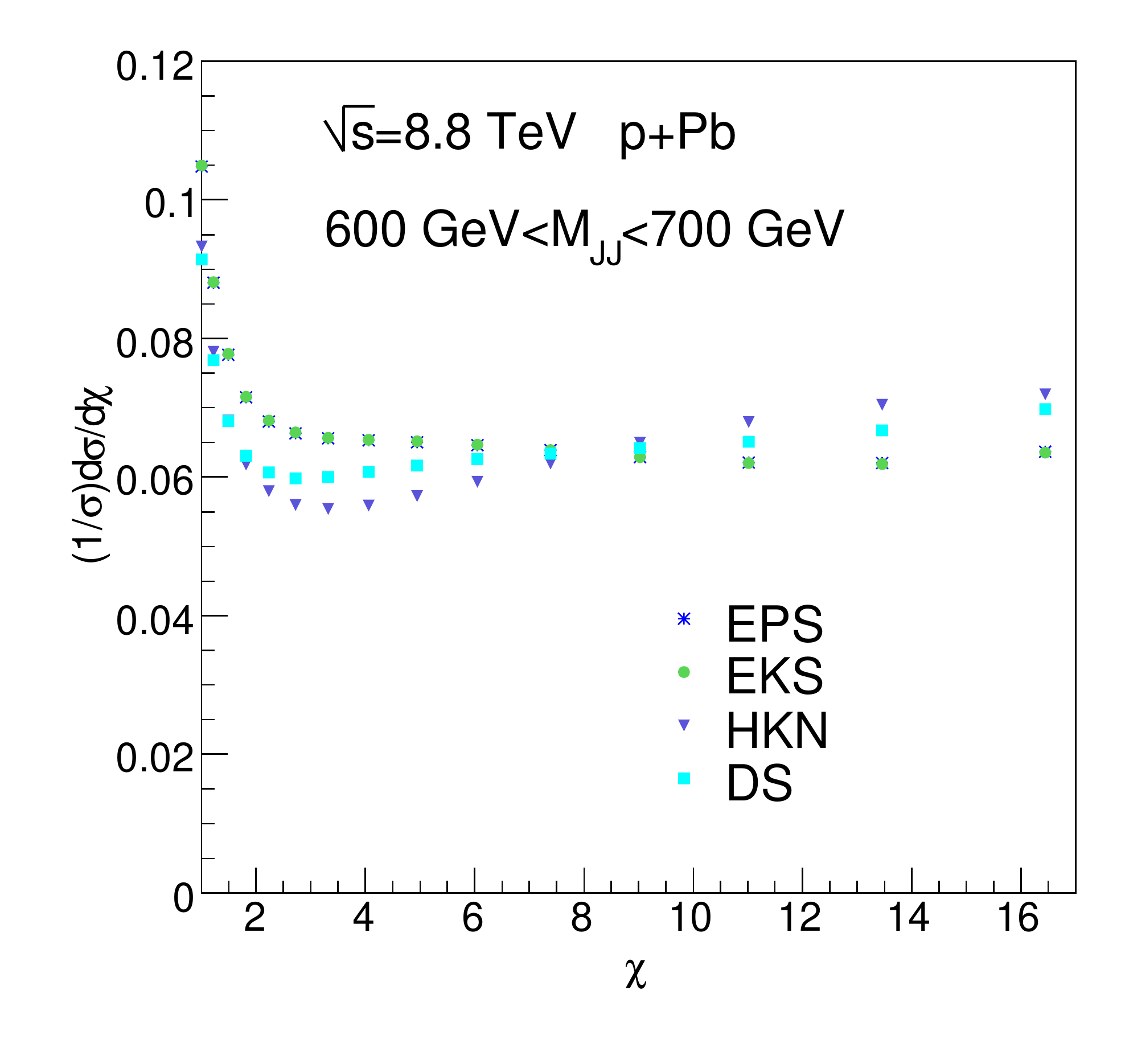}
\includegraphics[width=85mm]{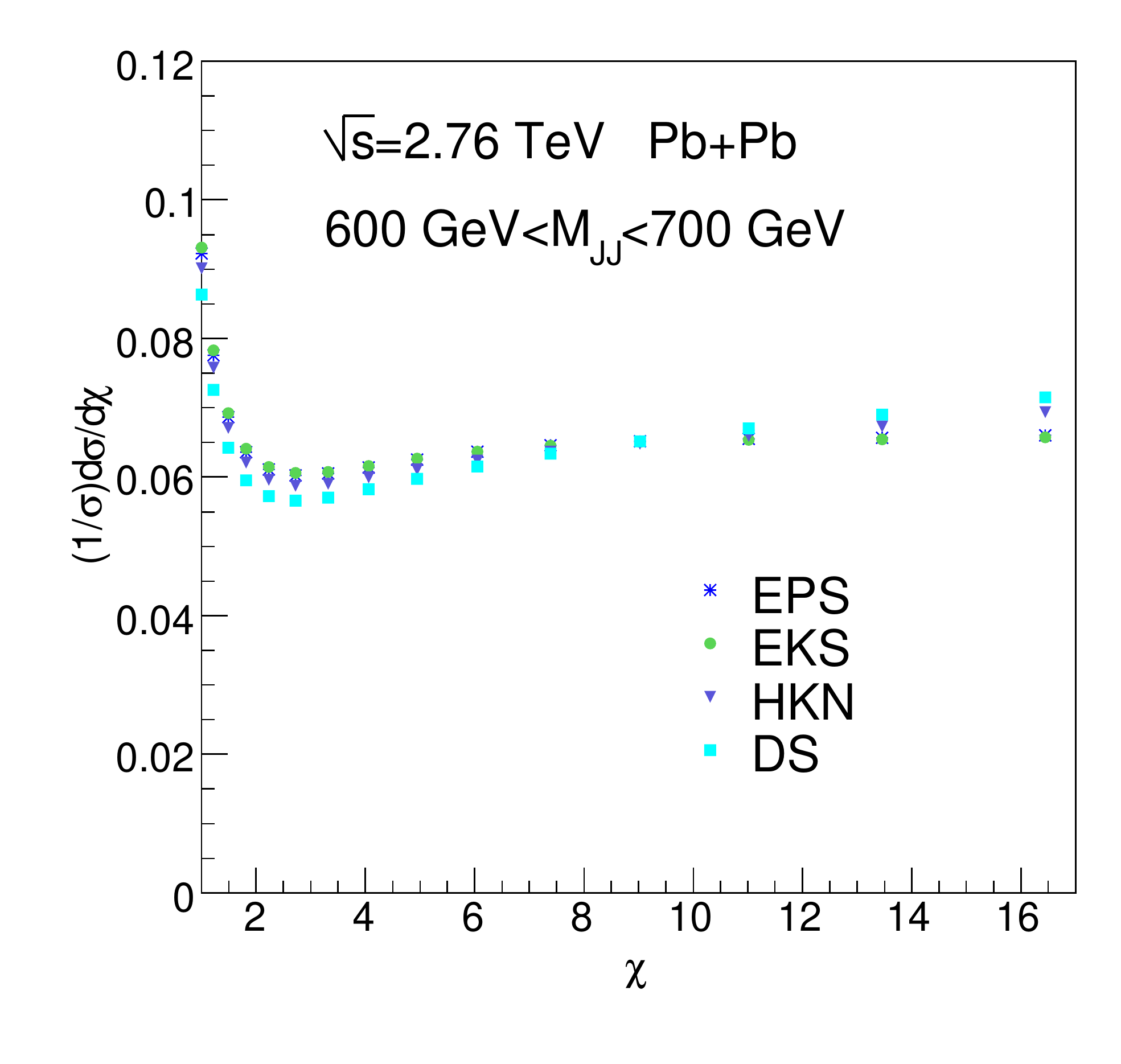}
    \caption{Dijet angular distributions at NLO are extended to p+Pb collisions at $\sqrt{s}=8.8$~TeV (top) and Pb+Pb collisions at $\sqrt{s}=2.76$~TeV (bottom) with the cone size $R=0.4$. Different sets of nPDFs given by EPS, EKS, HKN, DS are considered.}
    \label{dsigmadxtosig}
\end{figure}

In order to investigate the CNM effects on angular distributions in p+A and A+A collisions intuitively, we define
the nuclear modification factors of angular distributions,
\begin{eqnarray}
   R^{\chi}_{pA(AA)}=\frac{d\sigma_{pA(AA)}/\sigma_{pA(AA)} d\chi}{d\sigma_{pp}/\sigma_{pp} d\chi}
   \label{Rchi},
\end{eqnarray}
taking the  angular distribution in p+p collisions as a baseline.

Similarly, the nuclear modification factors for invariant mass spectrum $
d^{2}\sigma_{pA(AA)}
/dM_{JJ}d|y|_{max}$, final transverse momentum spectrum $d^{2}\sigma_{pA(AA)}/dE_{T1}dE_{T2}$, and momentum imbalance $d\sigma_{pA(AA)}/\sigma_{pA(AA)}dA_J$  can be written  respectively as
\begin{eqnarray}
  R^{M_{JJ}}_{pA(AA)}=\frac{d^{2}\sigma_{pA(AA)}/dM_{JJ}d|y|_{max}}{\langle N_{binary}\rangle d^{2}\sigma_{pp}/dM_{JJ}d|y|_{max}}
   \label{Rm},
\end{eqnarray}
\begin{eqnarray}
  R^{E_{T}}_{pA(AA)}=\frac{d^{2}\sigma_{pA(AA)}/dE_{T_1}dE_{T2}}{\langle N_{binary}\rangle d^{2}\sigma_{pp}/dE_{T1}dE_{T2}}
   \label{Re2},
\end{eqnarray}
\begin{eqnarray}
 R^{A_J}_{pA(AA)}=\frac{d\sigma_{pA(AA)}/\sigma_{pA(AA)}dA_J}{d\sigma_{pp}/\sigma_{pp}dA_J}
\label{Raj}.
\end{eqnarray}
The parameter $\langle N_{binary}\rangle$ in the denominator represents
 pairs of participating collisions. In p+A collisions it is decided by $\langle N_{binary}\rangle=\int d^{2} \mathbf{b}t_A(\mathbf{b})$. In A+A collisions it is $\langle N_{binary}\rangle=\int d^{2}\mathbf{b}\int d^{2}\mathbf{r}t_A(\mathbf{r})
 t_A(\mathbf{b}-\mathbf{r})$ .

\begin{figure}
\includegraphics[angle=90,width=85mm]{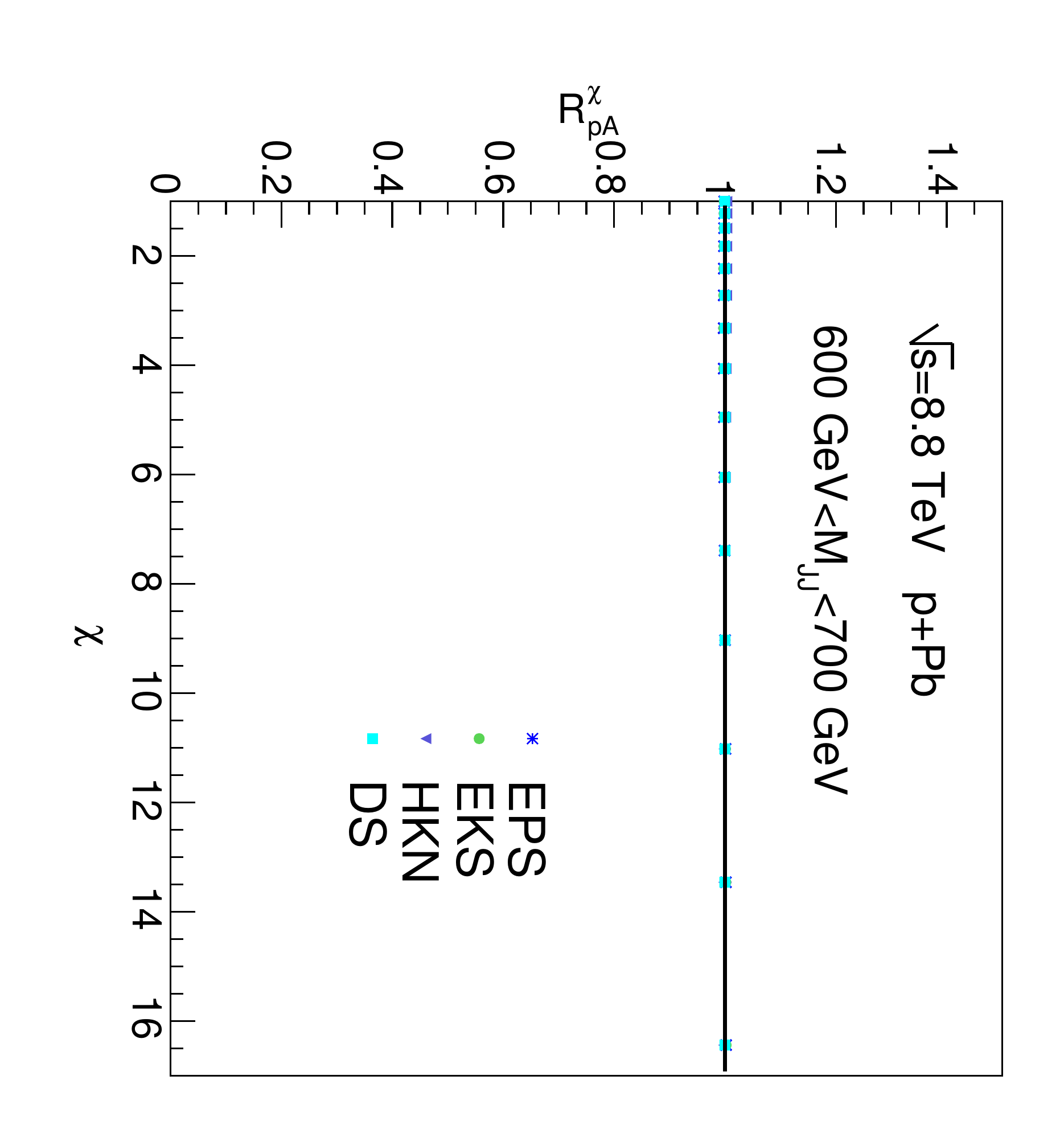}
\includegraphics[angle=90,width=85mm]{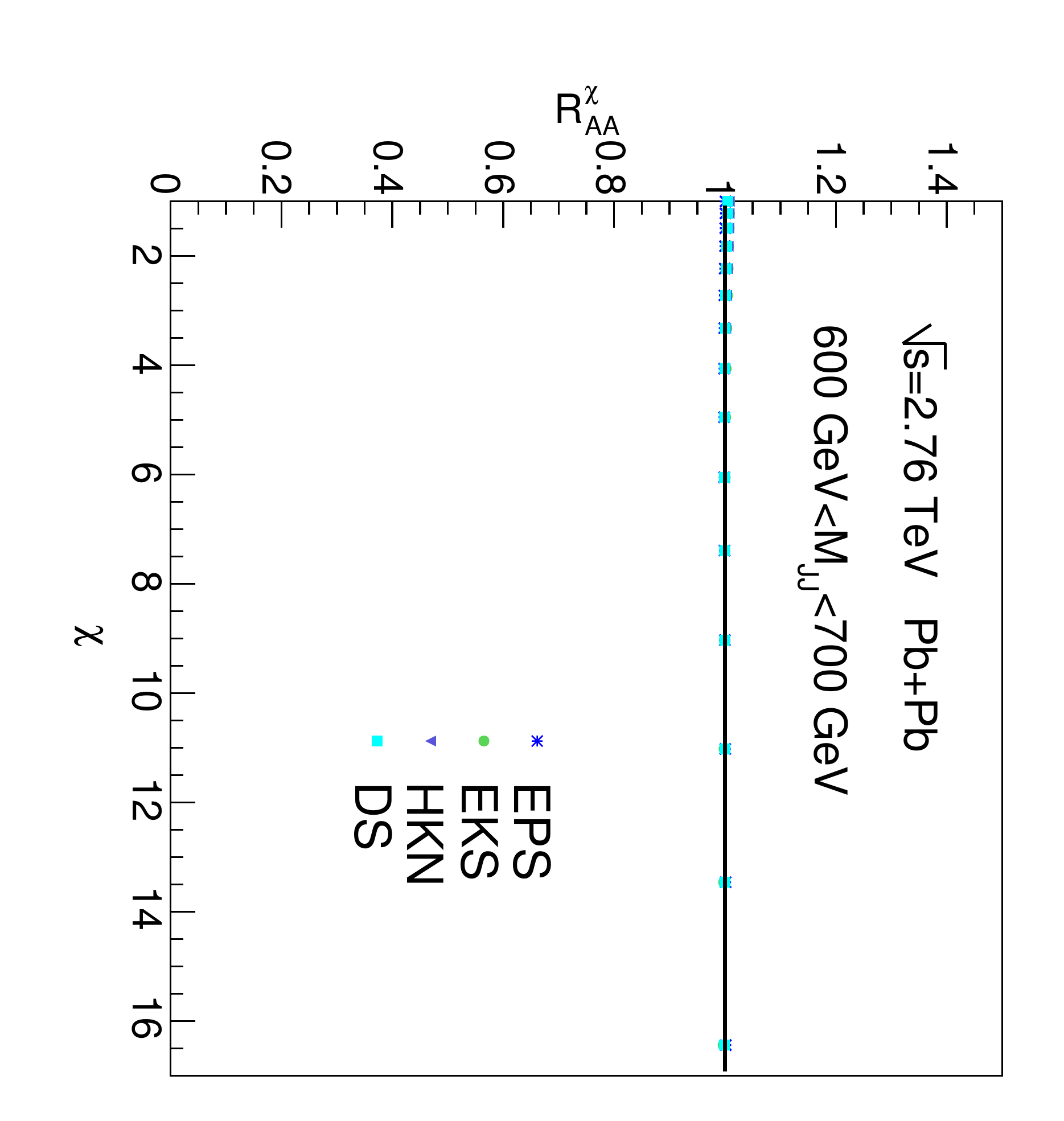}
    \caption{Nuclear modification factors of dijet angular distributions for p+Pb collisions at $\sqrt{s}=8.8$~TeV (top) and Pb+Pb collisions at $\sqrt{s}=2.76$~TeV (bottom) with different nPDFs.}
    \label{Rchipa}
\end{figure}

\begin{figure}
\includegraphics[width=85mm]{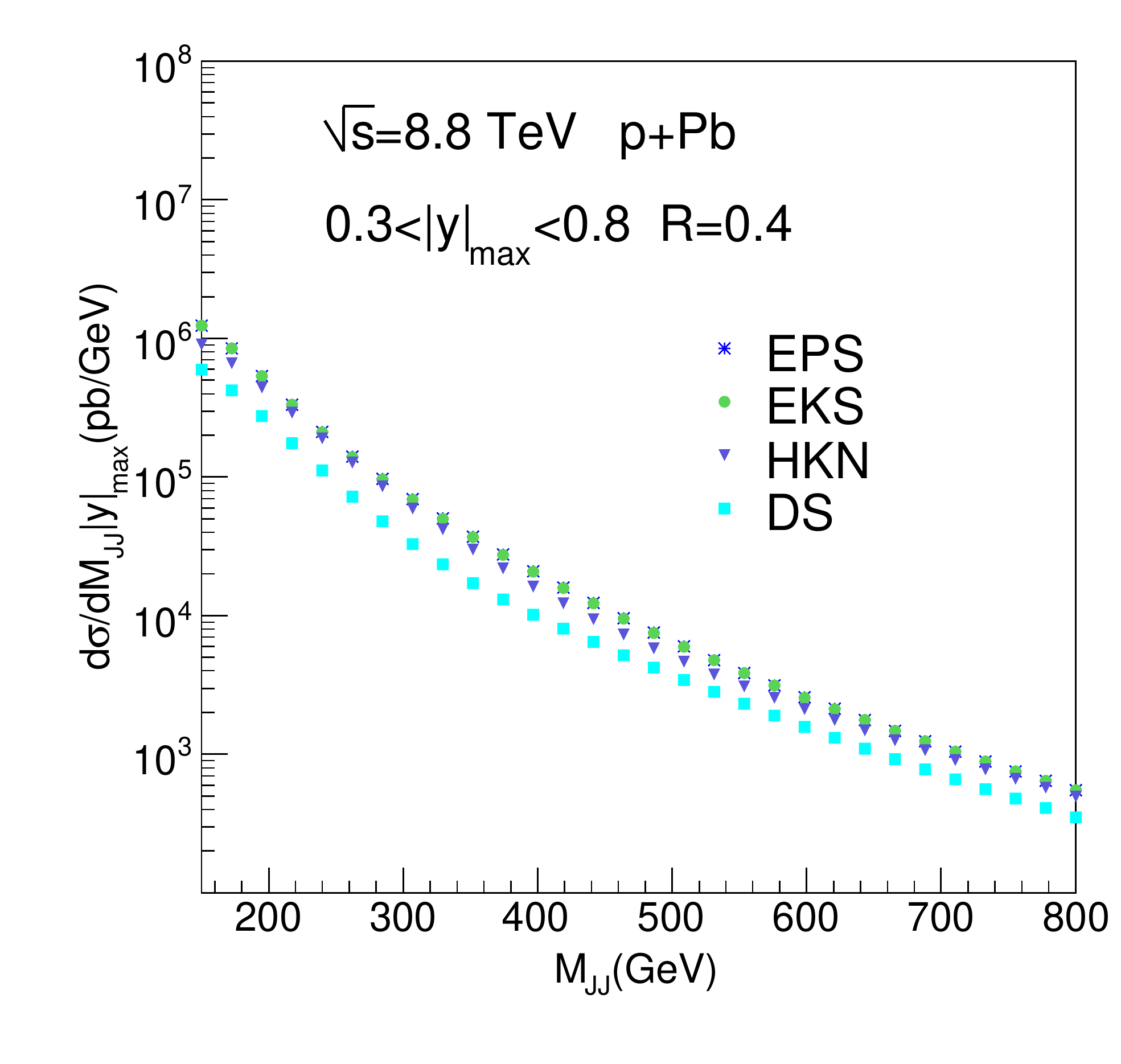}
\includegraphics[width=85mm]{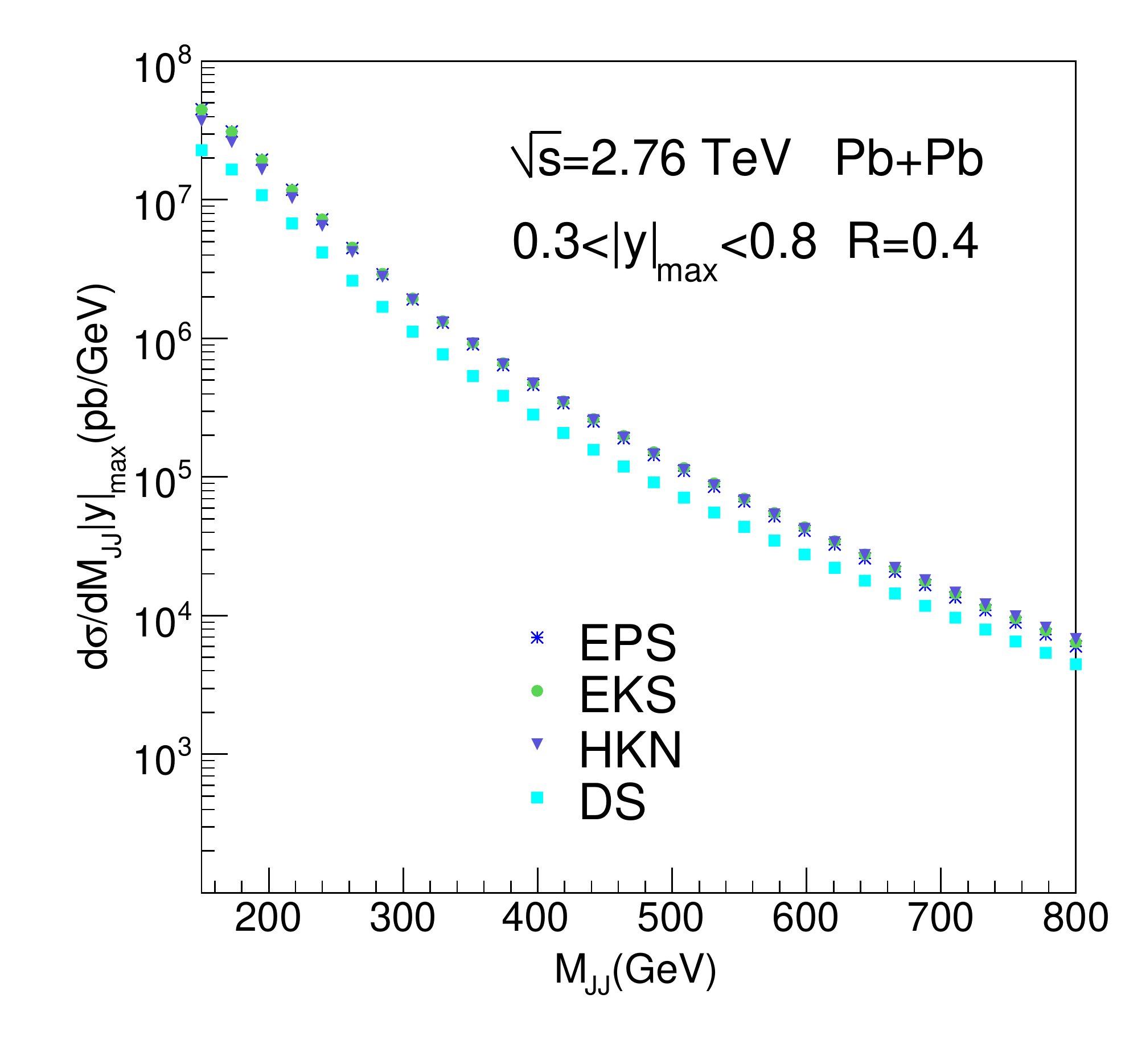}
    \caption{Dijet invariant mass spectra at NLO for p+Pb collisions at $\sqrt{s}=8.8$~TeV (top) and Pb+Pb collisions at $\sqrt{s}=2.76$~TeV(bottom) in the rapidity bin 0.3$<|y|_{max}<$0.8 with different nPDFs.}
    \label{dsigmjjpa aa}
\end{figure}
\begin{figure}
\includegraphics[width=85mm]{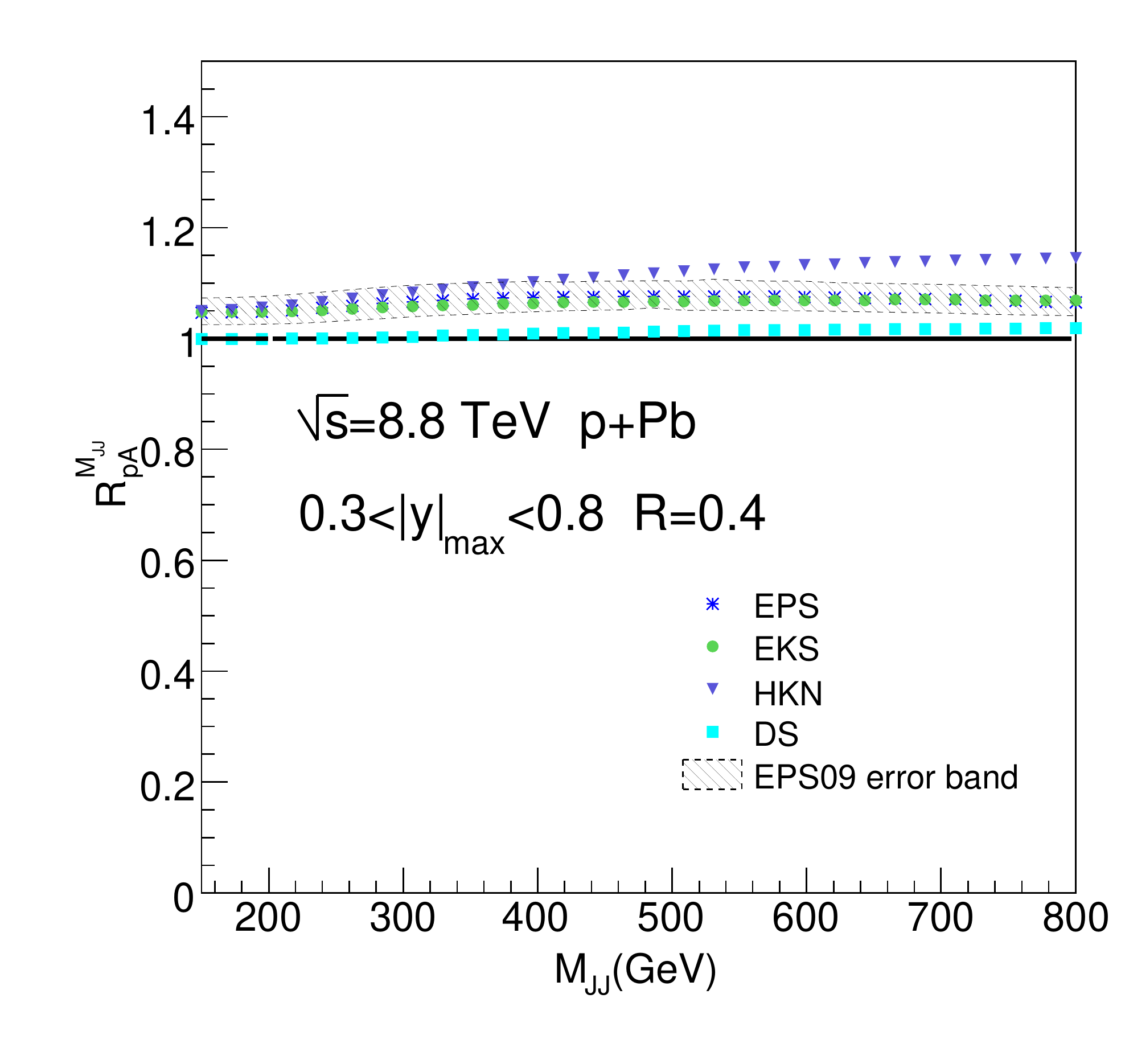}
\includegraphics[width=85mm]{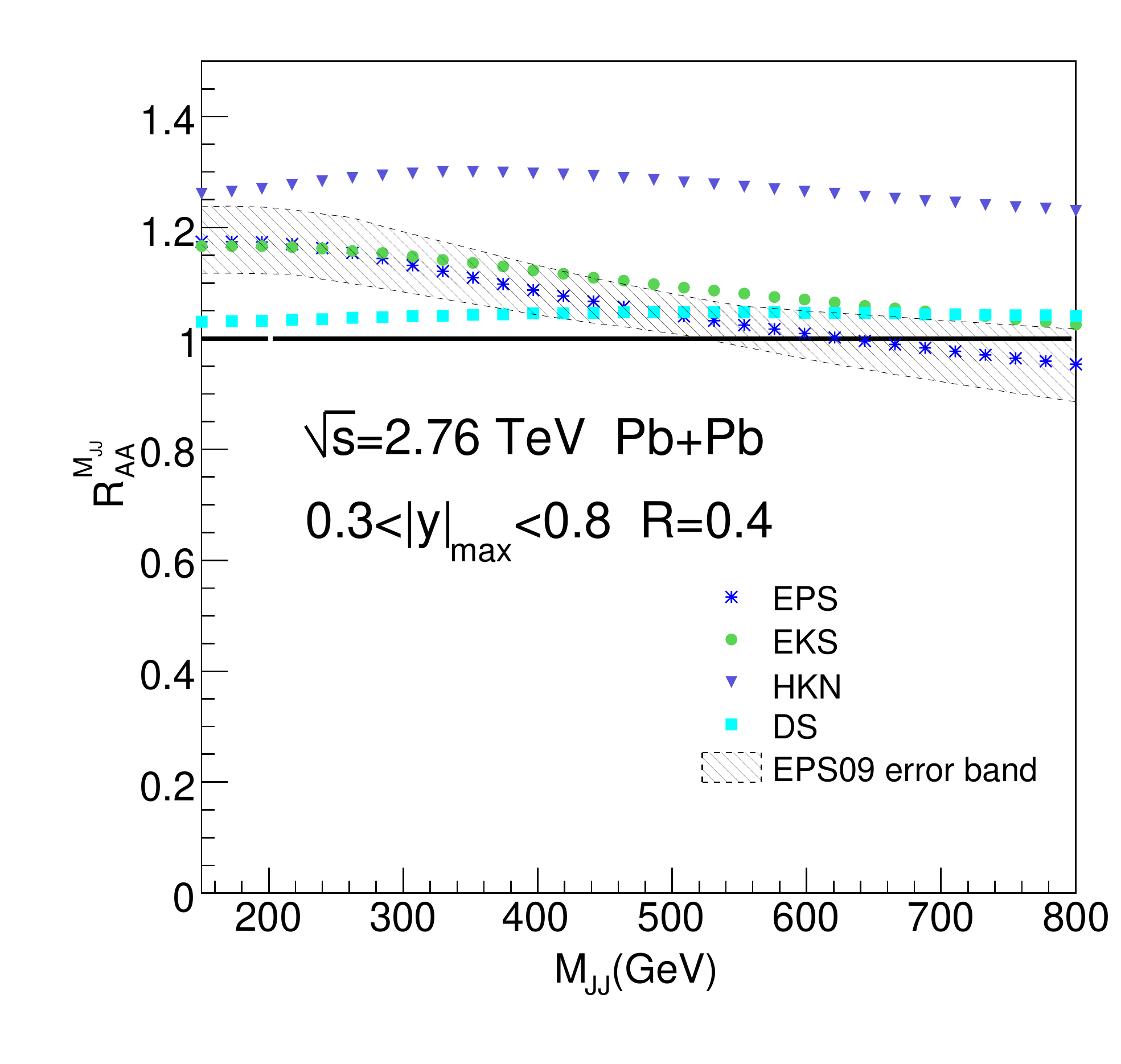}
    \caption{Nuclear modification factors of dijet invariant mass spectra for p+Pb collisions (top) and Pb+Pb collisions (bottom) with different nPDFs.}
    \label{Rm}
\end{figure}

Now we predict dijet productions in minimum bias p+Pb and Pb+Pb collisions at LHC due to the CNM effects. Please notice that for dijet
productions in Pb+Pb collisions the final-state hot QGP effects such as jet quenching in the hot QCD medium are not included in
the investigations presented in this paper, and in the following we focus only on the CNM effects on dijet productions in heavy-ion reactions.

Firstly we study the angular distributions in the mass interval of 600~GeV to 700~GeV and $|y_{JJ}|<1$
for p+Pb collisions at $\sqrt{s}=8.8$~TeV, for Pb+Pb collisions at $\sqrt{s}=2.76$~TeV.
 The jet radius is set to 0.4. Both angular
distributions in p+Pb and Pb+Pb shown in Figure~\ref{dsigmadxtosig} have identical trends with that in p+p
collisions. Results with different nPDF parameterizations EPS, EKS, HKN and DS,
show very small deviations.

 Figure~\ref{Rchipa} shows nuclear modifications of
 dijet angular distributions with different sets of nPDFs in p+Pb and Pb+Pb collisions at LHC.
 It reveals that the nuclear modifications are independent of the scattering angular $\chi$ and invisible
  over the whole range for both p+Pb and Pb+Pb collisions. This interesting feature results from the
  constrained kinematics regions probed by dijet angular distributions for fixed $M_{JJ}$. At LO from Eq.~(\ref{x}), Eq.~(\ref{y}) and Eq.~(\ref{M}), we can obtain $x_a*x_b=M_{JJ}^2/s$, which imposes a constraint for the momentum fractions $x_a$ and $x_b$. In particular when $y_{JJ}\sim 0$, at LO we find $x_a \sim x_b \sim M_{JJ}/\sqrt{s}$,
  which means that the momentum factions in nPDF show very weak dependence on the dijet angular $\chi$. We note that in A+A reactions the insensibility of dijet angular distribution  to the initial-state CNM effects implies the dijet angular distribution is a good tool to probe the final-state hot QCD matter effects such as jet quenching in the QGP.

   The invariant mass spectra in a rapidity bin $0.3<|y|_{max}<0.8$, with jet size $R=0.4$ for p+Pb collisions at $\sqrt{s}=8.8$~TeV, and Pb+Pb collisions at $\sqrt{s}=2.76$~TeV are plotted in Figure~\ref{dsigmjjpa aa}. The cross sections with four sets of nPDFs fall consistently with the dijet mass.

\begin{figure}
\includegraphics[width=85mm]{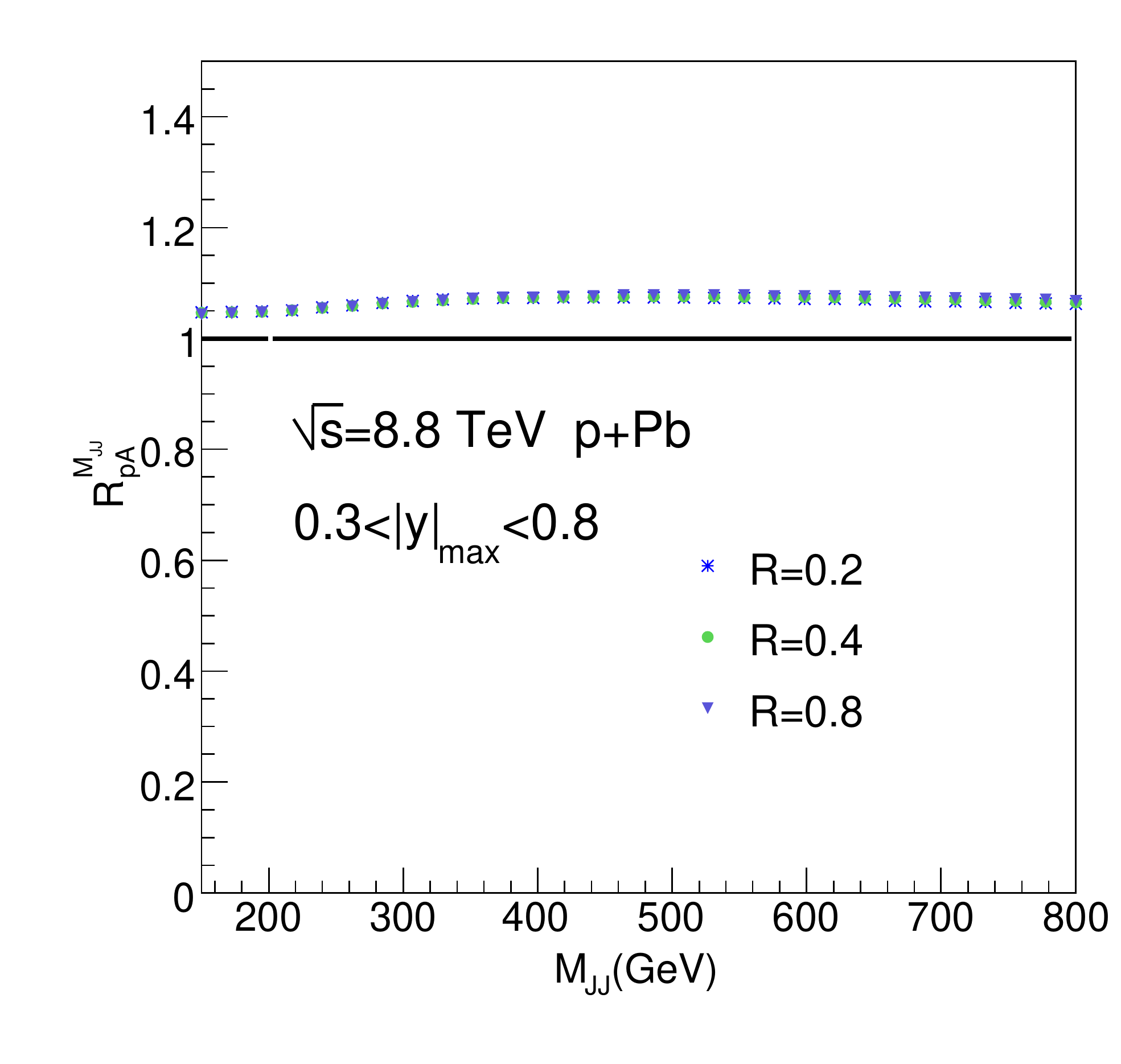}
\includegraphics[width=85mm]{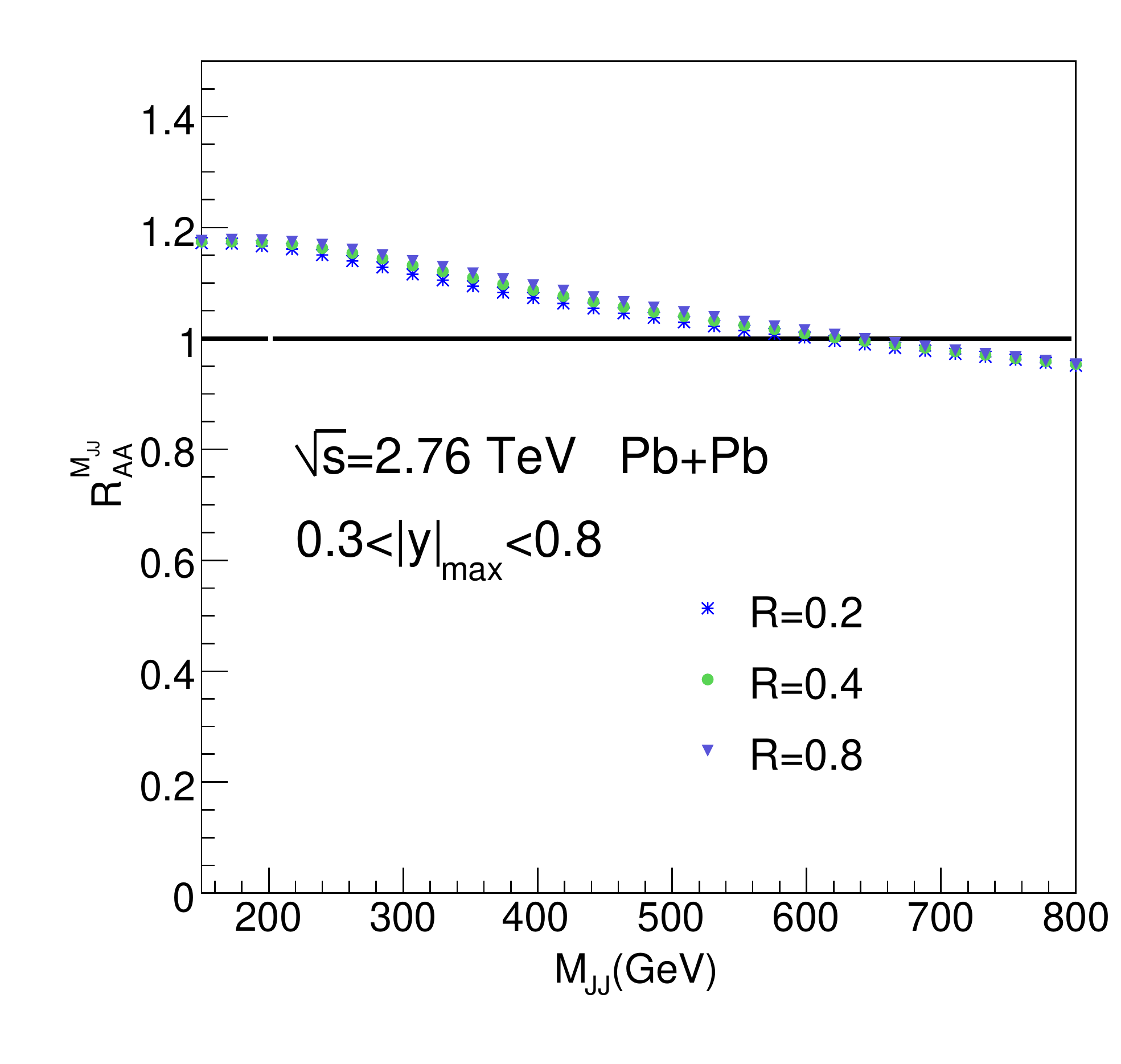}
    \caption{Nuclear modification factors of dijet invariant mass spectra for different jet cone sizes in a fixed rapidity bin 0.3$<|y|_{max}<$0.8. Top plane is the modification for p+Pb collisions and the bottom is for Pb+Pb collisions.}
    \label{Rm-3r}
\end{figure}

\begin{figure}
\includegraphics[width=85mm]{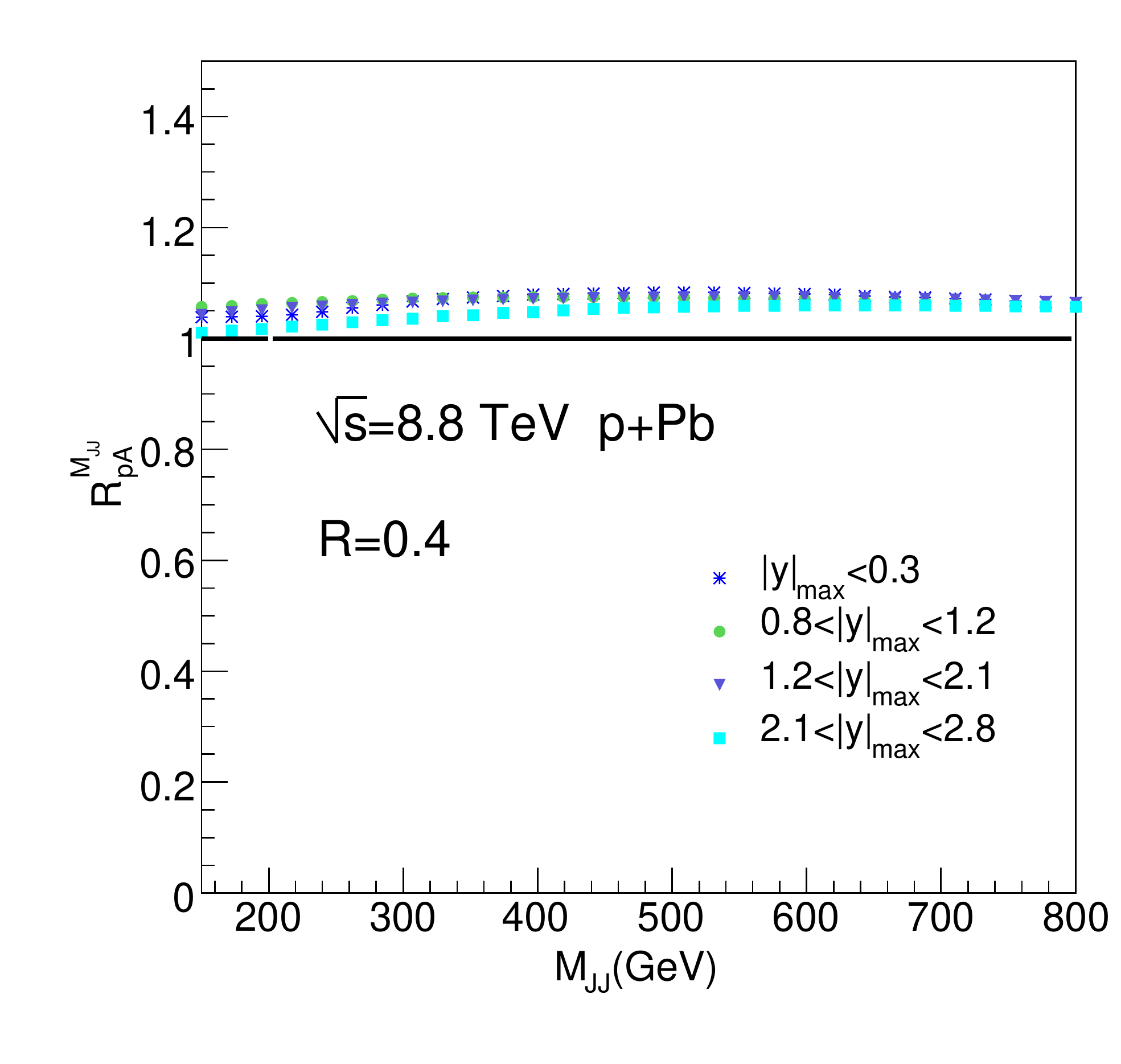}
\includegraphics[width=85mm]{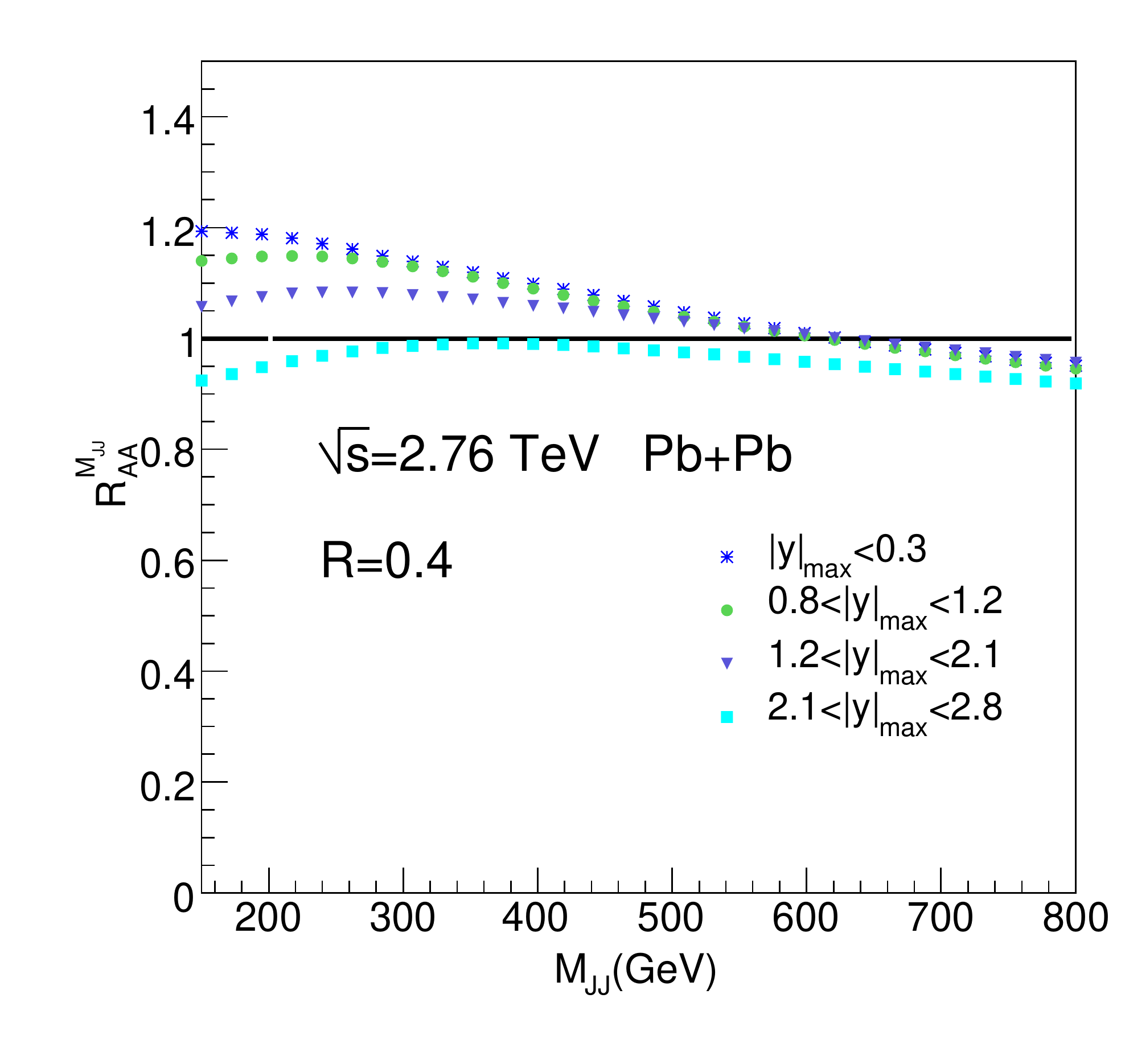}
    \caption{Nuclear modification factors of dijet invariant mass spectra for different rapidity bins with a fixed cone size $R$=0.4. Top plane is the modification for p+Pb collisions and the bottom is for Pb+Pb collisions.}
    \label{Rm-4y}
\end{figure}

Shown in Figure~\ref{Rm} are nuclear modification factors of dijet invariant mass spectra with 4 different sets of nPDFs. The overall trend is that the nuclear modification factors are greater than 1 due to the anti-shadowing effect in the kinematical region of $M_{JJ}$, except that results with
EPS  are slightly suppressed at very large $M_{JJ}$ in Pb+Pb reactions.  The top plane is the modification factor in p+Pb collisions at $\sqrt{s}=8.8$~TeV, and the CNM effects are not very visible. The bottom plane is the modification factor in Pb+Pb collisions at $\sqrt{s}=2.76$~TeV. In the mid-rapidity region, we can roughly estimate at LO $M_{JJ}\sim2E_{T}$ via Eq.~(\ref{M}), and $E_{T}\sim x\sqrt{s}/2$ from Eq.~(\ref{x}). Taking HKN set as an example, when $M_{JJ}$ lying above $\sim 150$~GeV at $\sqrt{s}=8.8$~TeV, and above $\sim 50~$GeV at $\sqrt{s}=2.76$~TeV the momentum fractions of the initial-state partons are in the anti-shadowing region. This leads the enhancement in the range shown in Figure~\ref{Rm}. There are deviations between different sets of nPDFs for both p+Pb and Pb+Pb collisions. Therefore the invariant mass spectrum of dijet may provide a convenient physical quantity to distinguish different parametrizations of nPDFs. Note that EPS09 provides the tool to study the theoretical uncertainty of the nPDFs parametrization
~\cite{Eskola:2009uj} and we calculate the resulting error band of  $R_{AB}^{M_{JJ}}$ on dijet invariant mass distribution
as illustrated in Figure~\ref{Rm} by utilizing 30 different parametrization sets of NLO nPDFs from EPS09. One can observe that
with the error band the deviations between results of EKS, EPS and DS are rather small whereas that of HKN
gives a considerably larger enhancement and shows a visible distinction with those of EKS, EPS and DS.

We further investigate the dependence of nuclear
modification for invariant mass spectrum  on the jet radius and rapidity bin with EPS nPDF.

Figure~\ref{Rm-3r} demonstrates the sensitivity of the modification factor for dijet invariant mass spectrum to the jet size. We check three jet cone sizes $R=0.2$, 0.4, 0.8 in a fixed rapidity bin $0.3<|y|_{max}<0.8$.
The modification factor in p+Pb collisions is almost independent on the jet cone radius $R$, while the modification factor in Pb+Pb collisions demonstrates a slight dependence on the jet radius. This phenomenon because of the CNM effects is opposite to the nuclear modifications for single jets~\cite{Vitev:2008rz} and $Z^0/\gamma ^*$-tagged jets~\cite{Neufeld:2010fj} with
final-state hot-dense medium effect in Pb+Pb collisions, where one observes strong dependence of $R^{AA}$ on the jet radius $R$ and jet quenching effect decreases with larger jet cone size because  more energy carried by radiated gluon will fall within the jet area when the jet radius $R$ becomes larger.
\begin{figure}
\begin{center}
\includegraphics[width=85mm]{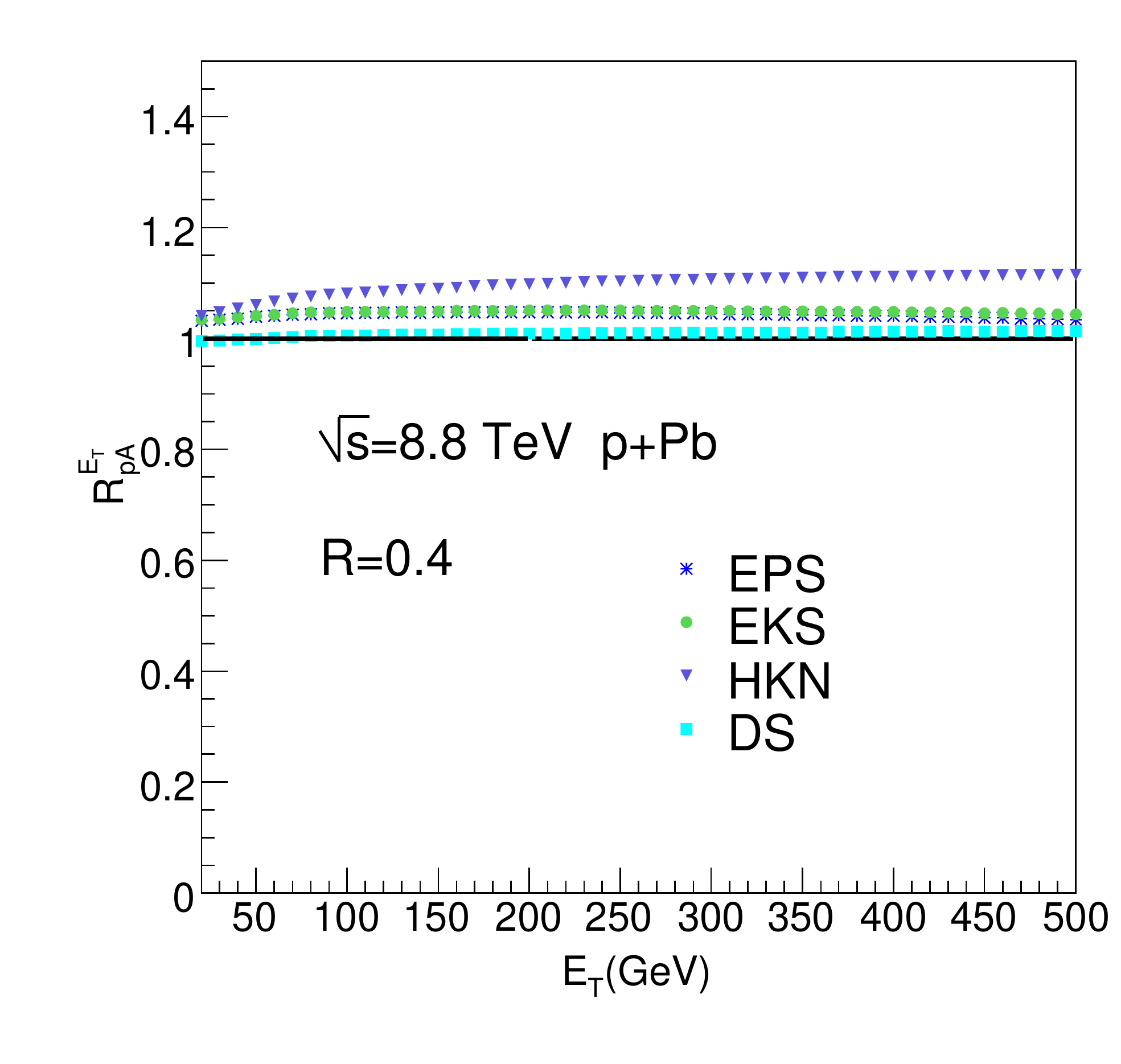}
\includegraphics[width=85mm]{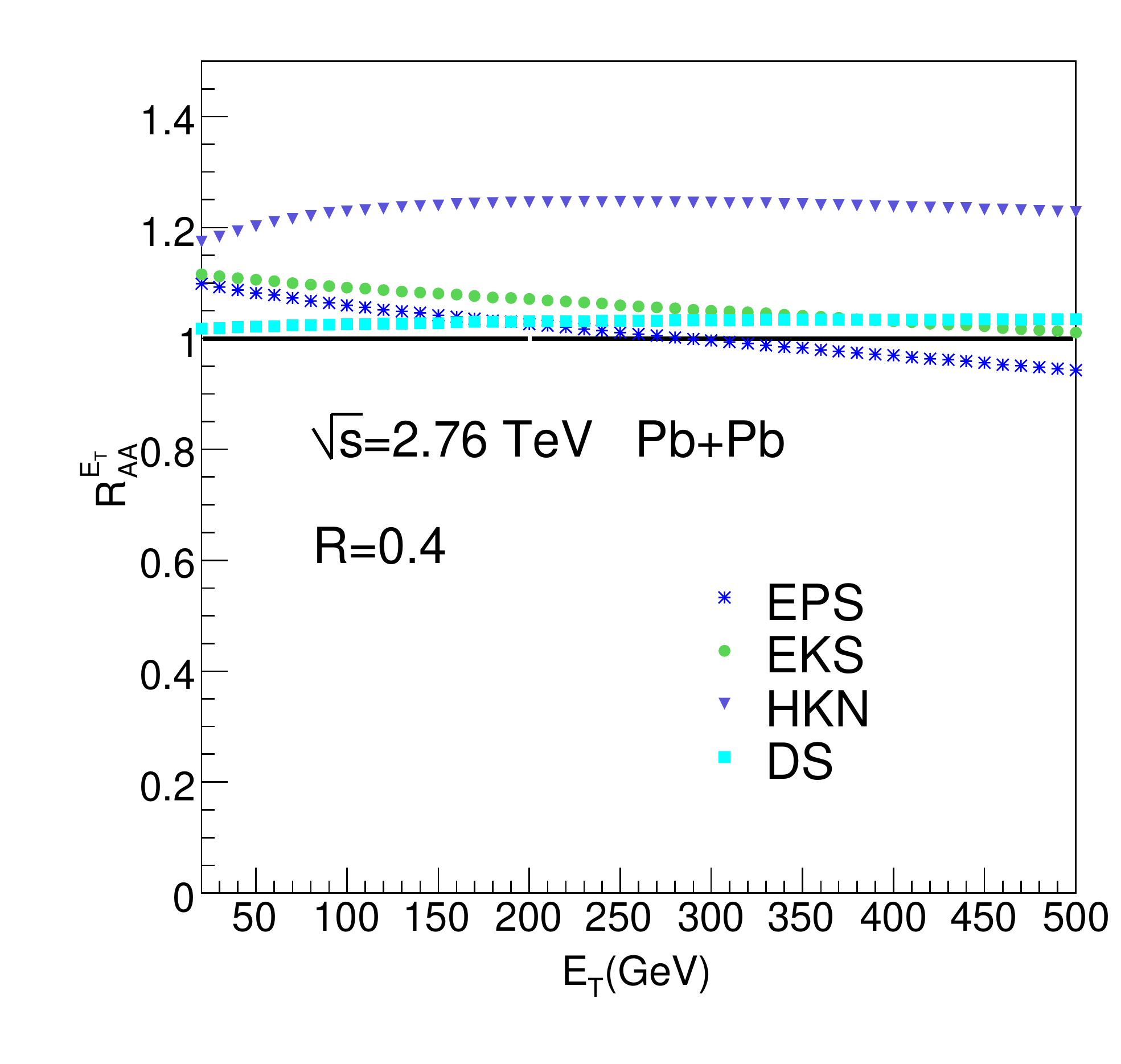}
\end{center}
    \caption{Nuclear modification factors for transverse energy spectra of dijet involving a fixed energy $E_{T1}=100$~GeV , in the set of different nPDFs.}
    \label{Re2}
\end{figure}

Figure~\ref{Rm-4y} illustrates the sensitivity of the modification factor for
dijet invariant mass spectrum to the rapidity bin. We vary rapidity in four
bins of $|y|_{max}<0.3$, $0.8<|y|_{max}<1.2$, $1.2<|y|_{max}<2.1$,
$2.1<|y|_{max}<2.8$ with the jet cone size $R=0.4$. The sensitivity of
nuclear modification in p+Pb collisions to the different rapidity bins is rather
modest,  and
we find that the modifications in central rapidity bins are stronger than that in 2.1$<|y|_{max}<$2.8. This kind of
dependence on rapidity region in Pb+Pb collisions becomes
apparent. The closer the rapidity bin is to central region, the
stronger the nuclear modification is, and for the largest rapidity region
$2.1<|y|_{max}<2.8$ a small suppression can be seen.
Final-state yield distributes in the central rapidity region dominantly, thus production in central rapidity is affected by nuclear matter effects most intensively.

\begin{figure}
\includegraphics[width=85mm]{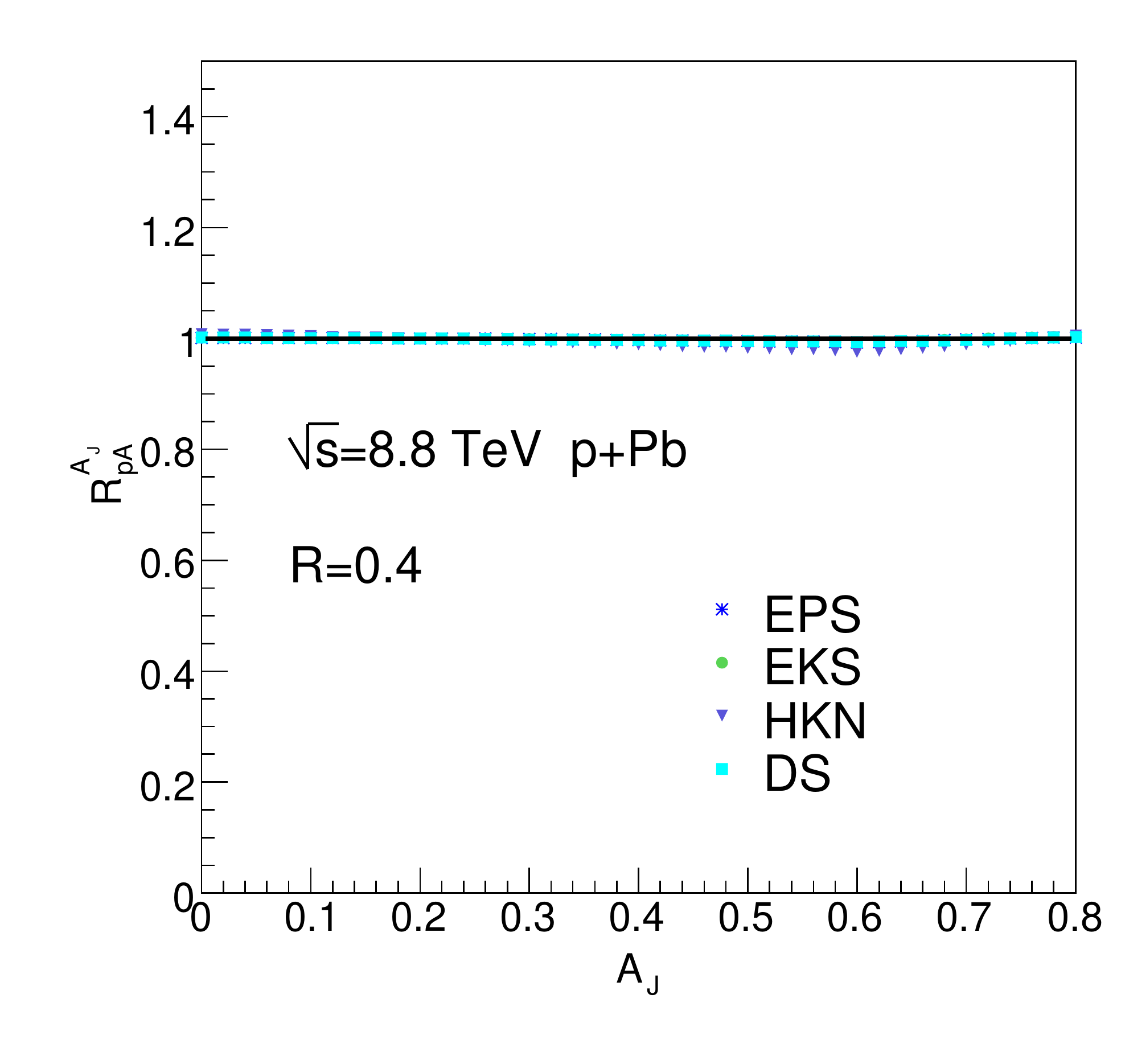}
\includegraphics[width=85mm]{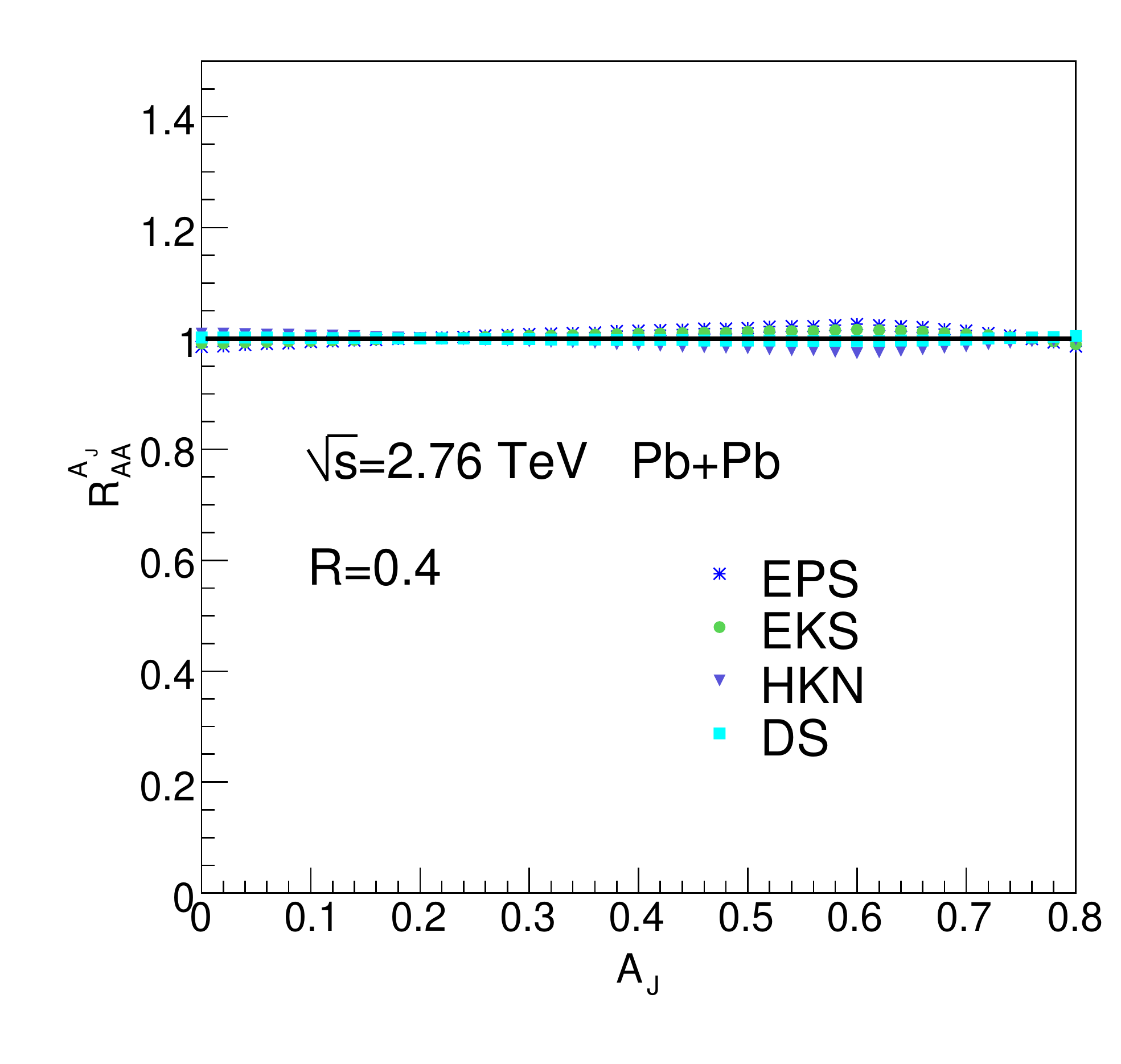}
    \caption{Nuclear modification factors of dijet asymmetry distribution for p+Pb collisions (top) and for Pb+Pb collisions (bottom) with different nPDFs.}
    \label{Raj}
\end{figure}

Fixing the transverse energy of a jet in final state $E_{T1}=100$~GeV, we survey the nuclear modifications for transverse momentum spectra of the other jet as displayed
in Figure ~\ref{Re2}. The yields in p+Pb and Pb+Pb collisions with different nPDFs are both enhanced, except that at very large $E_{T}$
results with EPS give a small suppression in Pb+Pb at $\sqrt{s}=2.76$~TeV, which are in accordance with the modifications
for dijet invariant mass spectra. The CNM effects increase the dijet yield in p+Pb collisions
about 10$\%$ at high $E_{T}$ and the yield in Pb+Pb collisions about20$\%$ at low $E_{T}$.
Furthermore, we can see the different modifications brought by different sets of nPDFs obviously. The difference
becomes slightly larger in Pb+Pb collision.

From transverse momentum spectra of dijet , we can derive the nuclear modifications of dijet asymmetry $A_J$ due to CNM effects with the cuts $E_{T1}>100$~GeV for leading jet and $E_{T2}>25$~GeV for subleading jet shown in Figure~\ref{Raj}.
We find that the nuclear modification on dijet asymmetry is similar with that on dijet angular distribution. The modification factors are close to unity and nearly independent of $A_J$ over the whole range. The fluctuation at large $A_J$ can be ignored since the yield at large $A_J$ is very small as revealed in Figure~\ref{daj}.The dependence of modification effects on $A_J$ is canceled by the fraction of transverse energy as in Eq.~(\ref{ajdef}). The CNM effects have little influence on dijet asymmetry. It shows that the rather flat $A_J$ distribution of dijet asymmetry measured by ATLAS and CMS in Pb+Pb collisions should be attributed to the final-state QGP effect. Thus the dijet asymmetry is an excellent signal to inspect the final-state QGP effects as well as dijet angular distribution due to their insensitivity to the CNM effects.

\section{Conclusion}
In high-energy nucleus-nucleus collisions particles productions at large
transverse momentum have been extended
from leading hadrons or vector bosons to reconstructed jets. Interesting
observations of dijet asymmetry in Pb+Pb
by ATLAS and CMS give a first successful example of jet measurements in
heavy-ion collisions and may demonstrate
the fingerprint of jet quenching in nuclear reactions at LHC for the
first time.  To put the study of dijet production in heavy-ion
collisions on a more reliable base the investigation of CNM effects is
indispensable.

In the paper we focus on CNM effects on the dijet productions in
relativistic heavy-ion collisions at LHC with the next-to-leading
order pQCD by incorporating different nPDFs set by EPS, EKS, HKN and DS. The dijet
angular distribution
$d\sigma_{pA(AA)}/\sigma_{pA(AA)}d\chi$, dijet invariant mass spectrum
$d^{2}\sigma_{pA(AA)}/dM_{JJ}d|y|_{max}$, final transverse energy spectrum of
 dijet $d^{2}\sigma_{pA(AA)}/dE_{T1}dE_{T2}$
 as well
as dijet momentum imbalance $d\sigma_{pA(AA)}/\sigma_{pA(AA)}dA_J$ at LHC are calculated
and the corresponding nuclear
modification factors due to CNM effects are given. We found that
dijet angular distributions and asymmetry distributions  are
insensitive to the
initial-state CNM effects and thus make themselves the excellent jet
observables to
investigate the final-state hot QGP effects such as jet quenching, while
the dijet invariant mass spectra and  transverse energy spectra
show  a conspicuous dependence on initial-state CNM effects.  The
overall enhancement of
$d^{2}\sigma_{pA(AA)}/dM_{JJ}d|y|_{max}$ over a wide region of $M_{JJ}$ and $d^{2}\sigma_{pA(AA)}/dE_{T1}dE_{T2}$
over a large region of $E_T$ for p+Pb and Pb+Pb at LHC
due to CNM effects is demonstrated, which is opposite to the
suppressions of jet spectra at high $E_T$
due to initial-state CNM effects at RHIC  and the
final-state jet quenching
effects both at RHIC and LHC. Visible difference between
utilizing of nPDFs of EPS, EKS, HKN and DS is
observed
in the nuclear modification factors for dijet invariant mass and transverse momentum spectra in
heavy-ion collisions, especially for large colliding system such as Pb+Pb, which
makes
$d^{2}\sigma_{pA(AA)}/dM_{JJ}d|y|_{max}$ and  $d^{2}\sigma_{pA(AA)}/dE_{T1}dE_{T2}$  potential observables
to distinguish different parametrizations of nPDFs. The sensitivity of
the nuclear modification of
invariant mass spectrum to jet radius $R$ and
rapidity $y$ is also explored. It is illustrated that though dependence
of $R^{M_{JJ}}$ on the jet size $R$ is negligible,
the variation of $R^{M_{JJ}}$  with different rapidity regimes is
relatively large in Pb+Pb collisions.

\begin{acknowledgements}
 This research is  supported
by the Ministry of Education of China with Project No. NCET-09-0411;
by NSFC of China with Project Nos. 11075062,
10825523 and 10875052; by MOST of China with Project No. 2008CB317106;
by NSF of Hubei Province with Project No. 2010CDA075;
and by MOE and SAFEA of China under Project No. PITDU-B08033.
\end{acknowledgements}


\begin{thebibliography}{9}
\bibitem{HP2010}
Talks and proceedings of Hard Probe 2010:
http://www.weizmann.ac.il/conferences/HP2010/ .

\bibitem{Wang:1991xy}
  X.~-N.~Wang, M.~Gyulassy,
  Phys.\ Rev.\ Lett.\  {\bf 68 } (1992)  1480-1483;
 M.~Gyulassy, I.~Vitev, X.~N.~Wang and B.~W.~Zhang,
 arXiv:nucl-th/0302077.


\bibitem{Baier:1996sk}
  R.~Baier, Y.~L.~Dokshitzer, A.~H.~Mueller, S.~Peigne, D.~Schiff,
  Nucl.\ Phys.\  {\bf B484 } (1997)  265-282;
  B.~G.~Zakharov,
  JETP Lett.\  {\bf 73 } (2001)  49-52;
  N.~Armesto, C.~A.~Salgado, U.~A.~Wiedemann,
  Phys.\ Rev.\  {\bf D69 } (2004)  114003.

\bibitem{Gyulassy:2000fs}
  M.~Gyulassy, P.~Levai, I.~Vitev,
  Phys.\ Rev.\ Lett.\  {\bf 85 } (2000)  5535-5538;
  I.~Vitev,
  Phys.\ Rev.\  {\bf C75 } (2007)  064906.

\bibitem{Wang:2001ifa}
 X.~N.~Wang and X.~F.~Guo,
  Nucl.\ Phys.\  A {\bf 696}, 788 (2001);
  B.~W.~Zhang and X.~N.~Wang,
  Nucl.\ Phys.\  A {\bf 720}, 429 (2003);
  B.~W.~Zhang, E.~k.~Wang and X.~N.~Wang,
  Nucl.\ Phys.\  A {\bf 757}, 493 (2005).





\bibitem{Borghini:2005em}
  N.~Borghini, U.~A.~Wiedemann,
  [hep-ph/0506218].

\bibitem{Lokhtin:2006dp}
  I.~P.~Lokhtin, S.~V.~Petrushanko, L.~I.~Sarycheva, A.~M.~Snigirev,
  Phys.\ Rev.\  {\bf C73}, 064905 (2006).

\bibitem{Vitev:2008rz}
  I.~Vitev, S.~Wicks and B.~W.~Zhang,
  JHEP {\bf 0811} (2008) 093

\bibitem{Vitev:2009rd}
  I.~Vitev and B.~W.~Zhang,
  Phys.\ Rev.\ Lett.\  {\bf 104} (2010) 132001

\bibitem{Neufeld:2010fj}
  R.~B.~Neufeld, I.~Vitev and B.~W.~Zhang,
  Phys.\ Rev.\  C {\bf 83} (2011) 034902



\bibitem{Aad:2010bu}
  G.~Aad {\it et al.} [ Atlas Collaboration ],
  Phys.\ Rev.\ Lett.\  {\bf 105}, 252303 (2010).

\bibitem{Chatrchyan:2011sx}
  S.~Chatrchyan {\it et al.}  [CMS Collaboration],
  Phys.\ Rev.\  C {\bf 84} (2011) 024906

\bibitem{Qin:2010mn}
  G.~Y.~Qin and B.~Muller,
  Phys.\ Rev.\ Lett.\  {\bf 106} (2011) 162302

\bibitem{Lokhtin:2011qq}
  I.~P.~Lokhtin, A.~V.~Belyaev and A.~M.~Snigirev,
  Eur.\ Phys.\ J.\  C {\bf 71} (2011) 1650




\bibitem{Young:2011qx}
  C.~Young, B.~Schenke, S.~Jeon, C.~Gale,
  [arXiv:1103.5769].

  \bibitem{He:2011pd}
  Y.~He, I.~Vitev, B.~-W.~Zhang,
  [arXiv:1105.2566].

\bibitem{CasalderreySolana:2010eh}
  J.~Casalderrey-Solana, J.~G.~Milhano, U.~A.~Wiedemann,
  J.\ Phys.\ G {\bf G38}, 035006 (2011).


\bibitem{Cacciari:2011tm}
  M.~Cacciari, G.~P.~Salam and G.~Soyez,
  Eur.\ Phys.\ J.\  C {\bf 71} (2011) 1692





\bibitem{Neufeld:2010dz}
R.~B.~Neufeld, I.~Vitev and B.~W.~Zhang,
  [arXiv:1010.3708].



\bibitem{Vitev:2008vk}
  I.~Vitev, B.~-W.~Zhang,
  Phys.\ Lett.\  {\bf B669}, 337-344 (2008).




\bibitem{Zhou:2010zzm}
  L.~-J.~Zhou, H.~Zhang, E.~Wang,
  J.\ Phys.\ G {\bf G37}, 105109 (2010).

\bibitem{Arleo:2011gc}
  F.~Arleo, K.~J.~Eskola, H.~Paukkunen, C.~A.~Salgado,
  JHEP {\bf 1104}, 055 (2011).

\bibitem{Xing:2011fb}
  H.~Xing, Y.~Guo, E.~Wang and X.~N.~Wang,
  [arXiv:1110.1903] .


\bibitem{Johnson:2000ph}
  M.~B.~Johnson {\it et al.} [ FNAL E772 Collaboration ],
  Phys.\ Rev.\ Lett.\  {\bf 86}, 4483-4487 (2001).


\bibitem{Ferreiro:2008wc}
  E.~G.~Ferreiro, F.~Fleuret, J.~P.~Lansberg and A.~Rakotozafindrabe,
  Phys.\ Lett.\  B {\bf 680}, 50 (2009);
R.~Vogt,
  Phys.\ Rev.\  C {\bf 81}, 044903 (2010).

\bibitem{Sharma:2009hn}
  R.~Sharma, I.~Vitev, B.~-W.~Zhang,
  Phys.\ Rev.\  {\bf C80}, 054902 (2009).

\bibitem{Duan:2011gu}
  C.~-G.~Duan, J.~-C.~Xu, L.~-H.~Song,
  [arXiv:1109.5337 [hep-ph]].




\bibitem{Zhang:2011ak}
  B.~-W.~Zhang,
  Nucl.\ Phys.\  {\bf A855}, 52-59 (2011).

\bibitem{Eskola:2009uj}
  K.~J.~Eskola, H.~Paukkunen, C.~A.~Salgado,
  JHEP {\bf 0904}, 065 (2009).


\bibitem{Frankfurt:2011cs}
  L.~Frankfurt, V.~Guzey, M.~Strikman,
  [arXiv:1106.2091 [hep-ph]].


\bibitem{Vitev:2007ve}
  I.~Vitev,
  Phys.\ Rev.\  {\bf C75}, 064906 (2007).





\bibitem{Eskola:1998df}
  K.~J.~Eskola, V.~J.~Kolhinen and C.~A.~Salgado,
  Eur.\ Phys.\ J.\  C {\bf 9} (1999) 61;
  K.~J.~Eskola, V.~J.~Kolhinen and P.~V.~Ruuskanen,
  Nucl.\ Phys.\  B {\bf 535} (1998) 351.


\bibitem{Hirai:2007sx}
  M.~Hirai, S.~Kumano and T.~H.~Nagai,
  Phys.\ Rev.\  C {\bf 76} (2007) 065207




\bibitem{deFlorian:2003qf}
  D.~de Florian and R.~Sassot,
  Phys.\ Rev.\  D {\bf 69} (2004) 074028.


\bibitem{Owens:1986mp}
  J.~F.~Owens,
  Rev.\ Mod.\ Phys.\  {\bf 59} (1987) 465.

\bibitem{Campbell:2006wx}
  J.~M.~Campbell, J.~W.~Huston and W.~J.~Stirling,
  Rept.\ Prog.\ Phys.\  {\bf 70} (2007) 89.

\bibitem{Ellis:2007ib}
  S.~D.~Ellis, J.~Huston, K.~Hatakeyama, P.~Loch and M.~Tonnesmann,
  Prog.\ Part.\ Nucl.\ Phys.\  {\bf 60}, 484 (2008)

\bibitem{Kunszt:1992tn}
  Z.~Kunszt and D.~E.~Soper,
  Phys.\ Rev.\  D {\bf 46} (1992) 192.
\bibitem{Ellis:1990ek}
  S.~D.~Ellis, Z.~Kunszt and D.~E.~Soper,
  Phys.\ Rev.\ Lett.\  {\bf 64} (1990) 2121.




\bibitem{theta}
  R.K. Ellis, W.J. Stirling, and B.R. Webber,
  QCD and Collider Physics (Cambridge University Press, Cambridge,
  England, 1996), p. 435, and references therein.

\bibitem{Eichten:1983hw}
  E.~Eichten, K.~D.~Lane and M.~E.~Peskin,
  Phys.\ Rev.\ Lett.\  {\bf 50}, 811 (1983).

\bibitem{Eichten:1984eu}
  E.~Eichten, I.~Hinchliffe, K.~D.~Lane and C.~Quigg,
  Rev.\ Mod.\ Phys.\  {\bf 56}, 579 (1984)
  [Addendum-ibid.\  {\bf 58}, 1065 (1986)].


\bibitem{Lane:1996gr}
  K.~D.~Lane,
  arXiv:hep-ph/9605257.

\bibitem{ArkaniHamed:1998rs}
  N.~Arkani-Hamed, S.~Dimopoulos and G.~R.~Dvali,
  Phys.\ Lett.\  B {\bf 429}, 263 (1998)

\bibitem{Atwood:1999qd}
  D.~Atwood, S.~Bar-Shalom and A.~Soni,
  Phys.\ Rev.\  D {\bf 62}, 056008 (2000)

\bibitem{Dienes:1998vg}
  K.~R.~Dienes, E.~Dudas and T.~Gherghetta,
  Nucl.\ Phys.\  B {\bf 537}, 47 (1999)

\bibitem{Pomarol:1998sd}
  A.~Pomarol and M.~Quiros,
  Phys.\ Lett.\  B {\bf 438}, 255 (1998)

\bibitem{Cheung:2001mq}
  K.~m.~Cheung and G.~L.~Landsberg,
  Phys.\ Rev.\  D {\bf 65}, 076003 (2002)



\bibitem{Pumplin:2002vw}
  J.~Pumplin, D.~R.~Stump, J.~Huston, H.~L.~Lai, P.~M.~Nadolsky, W.~K.~Tung,
  JHEP {\bf 0207}, 012 (2002).


\bibitem{:2009mh}
  V.~M.~Abazov {\it et al.}  [D0 Collaboration],
  Phys.\ Rev.\ Lett.\  {\bf 103} (2009) 191803.






\bibitem{Khachatryan:2011as}
  V.~Khachatryan {\it et al.}  [CMS Collaboration],
  Phys.\ Rev.\ Lett.\  {\bf 106} (2011) 201804


\bibitem{D0 PRD}
  B.~Abbott {\it et al.}  [D0 Collaboration],
   Phys.\ Rev.\  D {\bf 64}, 032003(2001).



\bibitem{:2010wv}
  G.~Aad {\it et al.}  [Atlas Collaboration],
  Eur.\ Phys.\ J.\  C {\bf 71} (2011) 1512







\end{thebibliography}
\end{document}